\documentclass[pra,reprint,aps,showkeys,showpacs,unsortedaddress,runinaddress,superscriptaddress]{revtex4-1}
\usepackage{amsmath,amssymb,epic,eepic}
\usepackage[latin1]{inputenc}
\usepackage[T1]{fontenc}
\usepackage[english]{babel}
\usepackage{graphicx}
\usepackage{hyperref,url}
\hypersetup{ 
    colorlinks, 
    citecolor=red, 
    filecolor=blue, 
    linkcolor=blue, 
    urlcolor=blue 
}
\usepackage{listings}
\usepackage{color}
\usepackage{textcomp}
\newcommand{\omp}{\omega_{+}}
\newcommand{\np}{n_{+}}
\newcommand{\omm}{\omega_{-}}
\newcommand{\nm}{n_{-}}
\newcommand{\Neff}{N\omega_{-}/\omega_{+}}

\begin{document}
\title{Quantum oscillations in ultracold Fermi gases~:\\realizations with 
rotating gases or artificial gauge fields}

\author{Ch.~Grenier} 
\affiliation{Centre de Physique Th\'eorique, Ecole Polytechnique, CNRS, 91128 Palaiseau Cedex, France.}

\author{C.~Kollath}
\affiliation{D\'epartement de Physique Th\'eorique, Universit\'e de Gen\`eve, CH-1211 Geneva, Switzerland.}

\author{A.~Georges}
\affiliation{Centre de Physique Th\'eorique, Ecole Polytechnique, CNRS, 91128 Palaiseau
Cedex, France.}
\affiliation{Coll\`ege de France, 11 place Marcelin Berthelot, 75005 Paris, France.}
\affiliation{DPMC-MaNEP, Universit\'e de Gen\`eve, CH-1211 Geneva, Switzerland.}

 
\begin{abstract}
We consider the angular momentum of a harmonically trapped, noninteracting Fermi gas subject to either 
rotation or to an artificial gauge field.  
The angular momentum of the gas is shown to display oscillations as a function of the particle number or chemical potential. This phenomenon is
analogous to the de Haas - van Alphen oscillations of the magnetization in the solid-state context.
However, key differences exist between the solid-state and ultracold atomic gases that we point
out and analyze. We explore the dependence of the visibility of these oscillations on the physical
parameters and propose two experimental protocols for their observation. Due to the very strong
dependence of the amplitude of the oscillations on temperature, we propose their use as
a sensitive thermometer for Fermi gases in the low temperature regime.
\end{abstract}


\maketitle

\section{Introduction}

When subject to an applied magnetic field, the electron gas of a metal displays an oscillatory dependence on
physical observables as a function of field strength. Generally known as 'quantum oscillations', these effects can
be observed for example for the magnetization (de Haas-
van Alphen oscillations), the resistivity (Shubnikov-de
Haas oscillations), the Hall resistance, or the specific
heat. Quantum oscillations are a macroscopic manifestation of the Landau level quantization of the spectrum
of electrons in the presence of a magnetic field.
They were predicted by Landau~\cite{Landau} and observed by de Haas and van Alphen~\cite{DHVA} in 1930. 
Since then,
quantum oscillations have become an invaluable tool in
characterizing the electronic states of metals and semi-
metals with large mean-free path. 
The measurement of the oscillations allows in particular for a
quantitative determination of Fermi surfaces and of the
effective masses of quasiparticles. A recent illustration is
the remarkable observation of quantum oscillations in underdoped cuprates in large magnetic fields~\cite{Naturecuprates}. This measurement has 
triggered many discussions about the reconstruction of the
Fermi surface in these systems and the possible relevance
of Fermi liquid descriptions.
In this article, we explore the possibility of observing
similar quantum oscillation effects for ultracold gases
of fermionic atoms, either by putting the gas in rotation or by using artificial gauge fields. We focus mostly
on the total angular momentum of the gas as a key observable, and identify the regimes in which oscillations
of this quantity as a function of e.g. atom number are
pronounced enough to be observable experimentally. We
emphasize the similarities, but also the key differences,
with the solid-state context.\\
Rotation has proven to be an efficient way to mimic the effect of a magnetic field for ultracold gases of 
neutral atoms. Over the last years, strong experimental and theoretical 
effort~\cite{RevModPhys.81.647,CooperReview,RevModPhys.80.885} has been devoted to the
understanding of rotating cold atomic gases. Experimental achievements include notably the observation of vortex nucleation in Bose-Einstein
condensates~\cite{PhysRevLett.85.2223} and Fermi gases~\cite{MZwNature}, which has been used as a proof of their superfluidity. On the theory side,
rotating gases are expected to be highly controllable test tables for strongly correlated phases of matter under the influence of a
magnetic field~\cite{RevModPhys.81.647,CooperReview,RevModPhys.80.885}, such as the fractional quantum Hall regime.
Recently, the experimental realization of artificial gauge fields with Raman beams~\cite{SPNature,RevModPhys.83.1523} opened
another promising way towards the generation of artificial magnetic fields for cold atoms.\\

In this paper, we analyze the influence of rotation or
of an artificial magnetic field on a noninteracting Fermi
gas, in a cylindrically symmetric harmonic trap. We focus mostly on the angular momentum of the gas with
respect to the rotation axis. The angular momentum
of a quantum gas differs from its classical counterpart.
While the former is defined from the change in energy
when varying rotation frequency, the latter is given by
the moment of inertia of the gas considered as a rigid
body and hence is related to its mean-square radius in
the plane perpendicular to the rotation axis. We compare
the quantum angular momentum to its classical counterpart by introducing their ratio. This quantity has already
been studied in the context of rotating Bose-Einstein condensates. 
 When Bose-Einstein condensation occurs, this ratio deviates from unity~\cite{PhysRevLett.76.1405}. 
It has also been shown to be proportional 
to the difference of frequencies of the $m=2$ quadrupolar oscillation modes~\cite{PhysRevLett.81.1754,PhysRevLett.86.2922,PhysRevLett.85.2223}.
In the case of superfluid helium, this ratio can give access to the ratio of the normal and superfluid fractions~\cite{PhysRevLett.25.1543,LeggetJSPHys}.

In the fermionic gas considered in this article, the ratio
of the quantum angular momentum to its classical counterpart tends to unity for very large atom number. We
show here that for intermediate values of the atom number and at low enough temperature, this ratio displays
pronounced quantum oscillations as a function of atom
number or chemical potential (Sec.~\ref{sec:rotation_framework}, Sec.~\ref{sec:slowrot}). Considering both the slow and fast rotation regime, we provide
(Sec.~\ref{sec:qualit_interp}) a simple qualitative picture of these oscillations
and emphasize the similarities but also key differences
with the de Haas-van Alphen oscillations in the solid-
state context~\cite{abrikosov1972introduction}. We perform a detailed analytical and
numerical analysis of the dependence of the amplitude
and period of these oscillations on temperature, rotation
frequency, atom number or chemical potential and aspect
ratio of the trap (Sec.~\ref{sec:qcontributionsL}). We emphasize in particular
that this oscillatory behavior, due to its sensitivity to
temperature, may be used as an accurate thermometer
in the low ($T /T_F < 0.1$) temperature regime. Finally,
we propose (Sec.~\ref{sec:Exp_imp}) two different experimental protocols that aim at measuring the angular momentum of
the gas, and thus to reveal its oscillating behavior with
respect to particle number.

\section{Model and observables}
\label{sec:rotation_framework}

  \subsection{General framework}

We consider a noninteracting Fermi gas in a cylindrically symmetric harmonic trap 
(with the radial and axial trapping frequencies being $\omega_0$ and $\omega_z$, respectively), rotating at an angular frequency 
$\omega$ around the symmetry axis $Oz$. The Hamiltonian in the rotating frame has the following form~\cite{RevModPhys.81.647,CooperReview,RevModPhys.80.885}:
\begin{equation}
\label{eq:1ph}
\mathcal{H} = \frac{\vec{p}^2}{2M}+\frac{1}{2}M\omega_0^2(x^2+y^2)+\frac{1}{2}M\omega_z^2z^2-L_z\omega\,.
\end{equation}
where $\vec{p}$ is the momentum, $M$ is the atomic mass and $L_z$ is the angular momentum with respect to the rotation axis.
The Hamiltonian \eqref{eq:1ph} is an analogue of the Hamiltonian describing an electron moving in a magnetic field in the solid state context. The angular momentum is the analogue of the magnetization, and the angular frequency plays the role of the magnetic field. However, note that typically no external trapping potential is present in the solid state context. We will see that this can lead to quite different behavior.  

The Hamiltonian \eqref{eq:1ph} gives rise to the following spectrum \cite{RevModPhys.81.647,Cohen}:
\begin{equation}
\label{eq:1ps}
 \varepsilon_{\nu} = \hbar\omega_+\left( n_+ + 1/2\right) + \hbar\omega_-\left( n_- + 1/2\right) + \hbar\omega_z\left( n_z + 1/2\right)\,,
\end{equation}
where $\omega_\pm = \omega_0 \pm \omega$ and $\nu = (n_+,n_-,n_z)$ with the nonnegative integers $n_+,n_-,n_z$. Interestingly, \eqref{eq:1ps} is identical to the spectrum of a noninteracting Fermi gas in a fully anisotropic harmonic trap,
with trapping frequencies $\omega_\pm$ and $\omega_z$.

The gas will be described at fixed inverse temperature $\beta=\frac{1}{k_BT}$, chemical potential $\mu$, and angular frequency $\omega$ in the grand canonical ensemble through the thermodynamic
potential $\Omega$ :
\begin{equation}
\label{eq:thpot}
\Omega \equiv \Omega (\beta,\mu,\alpha,\alpha_z) = -k_BT\sum_{\lbrace \nu \rbrace} \log{\left[ 1 + e^{\beta\left(\mu-\varepsilon_\nu\right)}\right]}\,,
\end{equation}
We defined the following dimensionless quantities~: 
\begin{equation}
\alpha = \frac{\omega}{\omega_0}\,\,,\text{and the aspect ratio of the trap}\,\,  \alpha_z = \frac{\omega_z}{\omega_0}\,,
\end{equation}
which together with $\frac{T}{T_F}$ will be the relevant parameters of the problem.\\

  \subsection{Angular momentum}

The angular momentum $\langle L_z \rangle$ of the gas is given  by the derivative of the thermodynamic potential
\eqref{eq:thpot} with respect to the angular frequency $\omega$~:
\begin{equation}
\label{eq:Ldef}
 \langle L_z \rangle \equiv -\frac{\partial \Omega}{\partial \omega}\,.
\end{equation}
Here, it reduces to:
\begin{align}
 \langle L_z \rangle & = \hbar \sum_{\lbrace \nu \rbrace} f(\varepsilon_\nu-\mu)(n_--n_+) \nonumber \\
{} & = \hbar \left( \langle n_- \rangle - \langle n_+ \rangle \right)
\label{eq:Lres}\,.
\end{align}
In the first equality, $f(\varepsilon) = \frac{1}{1+e^{\beta\varepsilon}}$ denotes the Fermi function at temperature $T$.\\

The angular momentum of a classical gas is proportional, in a rigid body-like motion, to the classical moment of inertia, which is proportional to its square 
extension around the rotation axis $z$ (cf. appendix \ref{sec:Rotation_classical})~:
\begin{equation}
\label{eq:Lclass_def}
 \langle L_z \rangle^{(cl)} = M\omega\langle x^2 + y^2 \rangle\,.
\end{equation}
It should be mentioned that this expression remains true even for interacting particles, as shown in appendix~\ref{sec:Rotation_classical}.

In order to derive a quantum analogue of the classical expression, we use that the square extension is given by the derivative of $\Omega$ (cf. eq.~\eqref{eq:thpot}) with respect to the squared radial trapping frequency $\omega_0^2$~:
\begin{equation}
 \langle x^2 + y^2 \rangle = -2\frac{\partial \Omega}{\partial (M\omega_0^2)}=\frac{\hbar}{M\omega_0}\sum_{\lbrace\nu\rbrace} f(\varepsilon_\nu-\mu)\left( n_+ + n_- + 1\right)
\end{equation}
This leads to the following expression for the quantum analogue of the classical angular momentum :
\begin{align}
 \langle L_z \rangle^{(cl)} & = \hbar\alpha\sum_{\lbrace\nu\rbrace} f(\varepsilon_\nu-\mu)\left( n_+ + n_- + 1\right)\\
		      {} & = \hbar\alpha\left(\langle n_+\rangle+\langle n_-\rangle + N\right)\,,
\label{eq:Lclass_result}
\end{align}
where $N = \sum_{\lbrace\nu\rbrace} f(\varepsilon_\nu-\mu)$ is the total number of particles of the gas.

In order to quantify the deviation of \eqref{eq:Lres} from its classical counterpart \eqref{eq:Lclass_result}, let us define the ratio
\begin{equation}
\label{eq:Ratiodef}
 R \equiv \frac{\langle L_z \rangle}{\langle L_z \rangle^{(cl)}} = \frac{\langle n_- \rangle - \langle n_+ \rangle
}{\alpha\left( \langle n_- \rangle + \langle n_+ \rangle + N\right)}\,,
\end{equation}
which depends on temperature, chemical potential (or equivalently on particle number), rotation, and 
axial confinement. In the limit of high temperature and large sample size, the gas behaves classically and the angular momentum is
well described by its classical expression \eqref{eq:Lclass_result}, and consequently $R=1$. Deviations from this classical analogue will be interpreted as manifestations of the
quantum behavior of the gas. The analysis of these deviations is the main subject of this work.


\section{Oscillations in the angular momentum}
\label{sec:slowrot}

\begin{figure}
 \centering
 \includegraphics[width=0.99\linewidth]{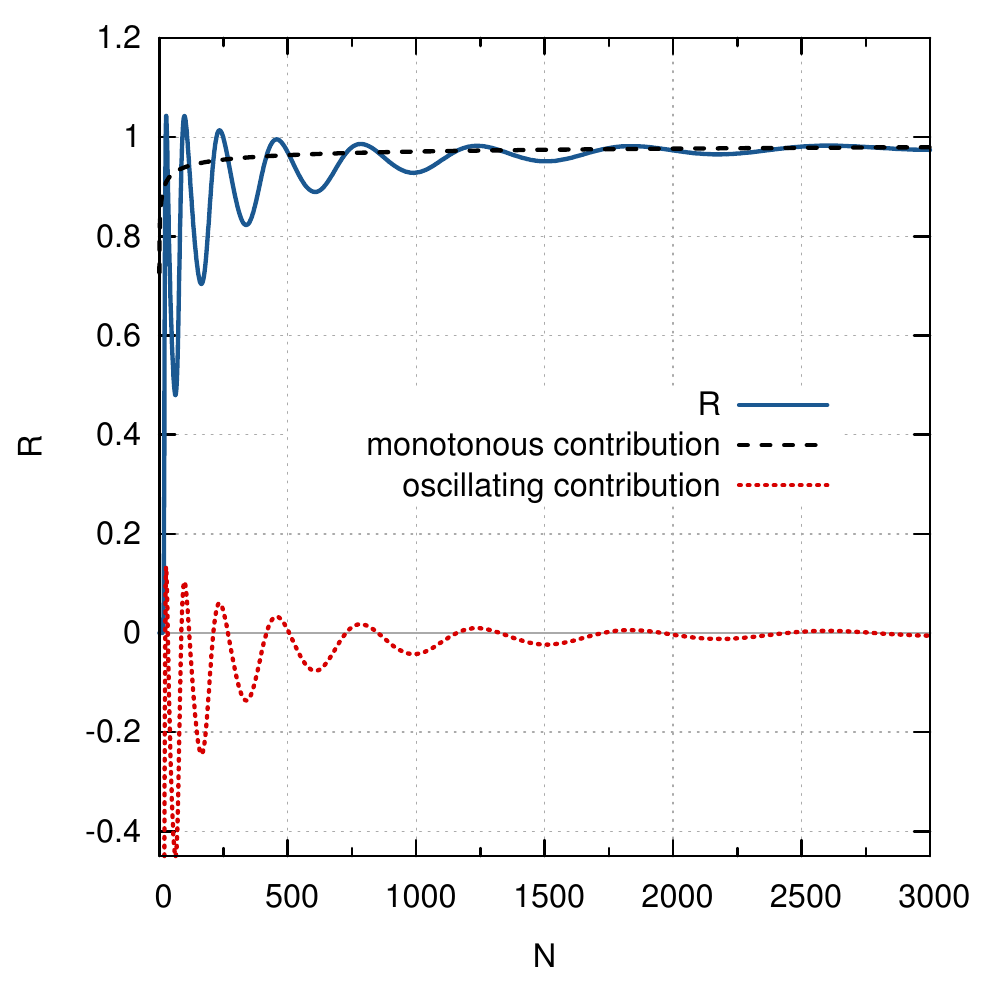}
 \caption{The ratio $R$, for $\alpha_z=0.05$, $\alpha=0.01$, and $T/T_F = 0.03$ (in blue). The black dashed line is the monotonous contribution, and
the red dotted line is the oscillating part. At high particle number, the oscillating contribution is negligible, and $R$ coincides with
its monotonous part~\eqref{eq:RNO}, as they both go to unity.}
\label{fig:example}
\end{figure}

Fig.~\ref{fig:example} depicts the evolution of $R$ with the atom number for a rotating fermion gas.

In this plot, one sees that the difference between the angular momentum of the gas~\eqref{eq:Ldef} and its classical analogue~\eqref{eq:Lclass_def} versus the particle number $N$ is twofold~:
\begin{itemize}
 \item A monotonous deviation $R_{mon}$, given in eq.~\eqref{eq:RNO} \footnote{This monotonous
deviation is also thought to be rather a quantum effect than a finite size one, since the classical expression for the angular momentum remains true even
for a single classical particle}
 \item An oscillating contribution $R_{osc}$ with $N$ which is very pronounced at low particle number\,.
\end{itemize}

In order to understand these two different contributions, we step back and investigate the thermodynamic potential $\Omega$ \eqref{eq:thpot} from which the physical quantities will be extracted.  
Appendix~\ref{sec:Psums} shows how the expression~\eqref{eq:thpot}, via Poisson summation, splits into two parts $\Omega = \Omega^{(no)}+\Omega^{(osc)}$. The first one denotes the non oscillating contribution with the chemical potential and is given by~:
\begin{equation}
 \Omega^{(no)} =-\frac{(k_BT)^4}{\hbar\omega_+\hbar\omega_-\hbar\omega_z}\text{Li}_4[-e^{\beta(\mu-\varepsilon_0)}].
\end{equation}
The second part is responsible for the oscillations with the chemical potential and is determined by 
\begin{align}
 \Omega^{(osc)} &=-2k_BT\Re{\left\lbrace\sum_{\vec{k}\neq\vec{0}} \widetilde{\Omega}_{\vec{k}}\right\rbrace},\,\,\text{with}\\
 \tilde{\Omega}_{\vec{k}}&=\int_{[0,+\infty[^3}d\vec{u} e^{2i\pi\vec{k}\cdot\vec{u}}\log{\left[1+e^{\beta(\mu-\varepsilon_{\vec{u}})}\right]}.
\end{align}
 In the previous equations,
$\text{Li}_n$ is the n-th polylogarithm function~\cite{AStegun}, $\vec{k}=(k_+,k_-,k_z)$ is the vector conjugated to 
$\vec{u}=(u_+,u_-,u_z)$ with $k_+$, $k_-$ and $k_z$ positive integers, $\varepsilon_0 = \hbar\omega_0 + \frac{1}{2}\hbar\omega_z$ is the ground
state energy and $\varepsilon_{\vec{u}} = \varepsilon_0 + \hbar\omega_+ u_+ + \hbar\omega_- u_- + \hbar\omega_z u_z$.

The splitting of the thermodynamic potential into an oscillating and a monotonous contribution with $\mu$ carries over to many observables. Typically, these oscillations stem from 'shell filling' effects as we will show in the next section. They have been predicted for different observables as e.g.~the radius 
of the cloud \cite{PhysRevA.57.1253,PhysRevLett.85.4648,PhysRevA.67.053601,PhysRevA.58.2427,PhysRevA.55.4346}. However, the relative strength of both contributions can vary very much with the observable considered. We will show that the oscillating contribution is much more pronounced in the angular momentum $\langle L_z\rangle$ of the gas with respect to the rotation axis then in 
previously considered observables \cite{PhysRevA.57.1253,PhysRevLett.85.4648,PhysRevA.67.053601,PhysRevA.58.2427,PhysRevA.55.4346}.\\

From the splitting of the thermodynamic potential into the two contributions, the analogue separation of both $\langle L_z \rangle$ and its classical analogue follows~:
\begin{equation}
 R = \frac{\langle L_z\rangle^{(no)}+\langle L_z\rangle^{(osc)}}{\langle L_z\rangle^{(cl,no)}+\langle L_z\rangle^{(cl,osc)}} = R_{mon}+R_{osc}\,.
\end{equation}

The monotonous contribution is given by~:
\begin{align}
\label{eq:Rmon}
 R_{mon} & = \frac{\langle L_z\rangle^{(no)}}{\langle L_z\rangle^{(cl,no)}}\\
{}& = \frac{2\text{Li}_4[-e^{\beta(\mu-\varepsilon_0)}]}{2\text{Li}_4[-e^{\beta(\mu-\varepsilon_0)}]+
(\beta\hbar\omega_0)^4(1-\alpha^2)^2\alpha_zN}\,.
 \label{eq:RNO}
\end{align}
The expression~\eqref{eq:RNO} goes to unity for large samples as depicted in Fig.~\ref{fig:example} which means that the classical value of the angular momentum is reached. The variations of the ratio $R$ with respect to 
$N$, $\alpha$, and $\alpha_z$ stem both from a direct dependence on these parameters as evident in~\eqref{eq:RNO}, but also from the indirect dependence via the chemical potential. For example, at zero temperature the chemical potential can be expressed by $\mu(T=0K) = \hbar\omega_0\left(6(1-\alpha^2)\alpha_zN\right)^{1/3}$. Using this expression 
and developing the $\text{Li}_n$ functions in~\eqref{eq:RNO}, one sees that
the strongest dependence is seen at low temperatures, where the expression simplifies to:
\begin{equation}
 R_{mon} = \frac{1}{1+2(1-\alpha^2)^{2/3}(6\alpha_z N)^{-1/3}}\,.
 \label{eq:RNOLowT}
\end{equation}

The oscillating part of $R$ (the red dotted line in Fig.~\ref{fig:example}) is recovered by expanding the denominator in~\eqref{eq:Rmon} in power series.
It is of the general form :  
\begin{align}
\label{eq:Rosc}
 R_{osc} = 2\Re\left\lbrace \sum_{\vec{k}\neq\vec{0}}\widetilde{R}(\vec{k})e^{2i\pi\mu\left(\frac{k_+}{\hbar\omega_+}+\frac{k_-}{\hbar\omega_-}+\frac{k_z}{\hbar\omega_z}\right)}\right\rbrace\,,
\end{align}
where $\widetilde{R}(\vec{k})$ contains contributions from $\langle L_z \rangle^{(no)}$, $\langle L_z \rangle^{(osc)}$, $\langle L_z \rangle^{(cl,no)}$,
$\langle L_z \rangle^{(cl,osc)}$. 

The expression~\eqref{eq:Rosc} states clearly that this contribution has a periodic dependence on the chemical potential, and therefore on the particle number, and that the characteristic
frequencies, that are the oscillation periods, are $\omega_\pm$ and $\omega_z$. The next section will shed some light on this oscillating
behavior, on the basis of qualitative arguments.


\section{Qualitative interpretation}
\label{sec:qualit_interp}

This section provides an intuitive interpretation of the oscillatory behavior of the angular momentum at low temperature. 
We shall first review a simple physical picture of quantum oscillations in a two-dimensional electron gas, and then point out the differences and similarities in the 
case of rotating cold atomic gases. We derive in particular simple analytical expressions for the angular momentum in the case 
of a two-dimensional system ($\omega_z \rightarrow \infty$) in the regime of fast rotation $\omega \lesssim \omega_0$. 

\subsection{Quantum oscillations in a two-dimensional electron gas: a simple picture}

Let us first consider the case of electrons  in a strictly two-dimensional electron gas (2DEG). Quantum oscillations have been 
reported in such systems, see e.g. Ref.~~\cite{2DEGQO}. In this case, such effects can be understood from a simple 
calculation which we now recall ~\cite{abrikosov1972introduction}. 

In the presence of a magnetic field $B$, the energy levels of the 2DEG are Landau levels with single-particle energies 
$\varepsilon_n=(n+1/2)\hbar\omega_c$, with $n\geq 0$ an integer and 
$\omega_c = \frac{eB}{m}$ the cyclotron frequency ($m$ is the mass of the electron and $-e$ its charge).
Each Landau level is macroscopically degenerate, with  the degeneracy
given by $\frac{\Phi}{\Phi_0}$, where $\Phi = BS$ is the magnetic flux over the whole sample, with $B$ the amplitude of the 
magnetic field and $S$ the surface of the sample, and $\Phi_0 = \frac{h}{e}$ is the 
flux quantum. 

\begin{figure}
 \centering
 \includegraphics[width=0.99\linewidth]{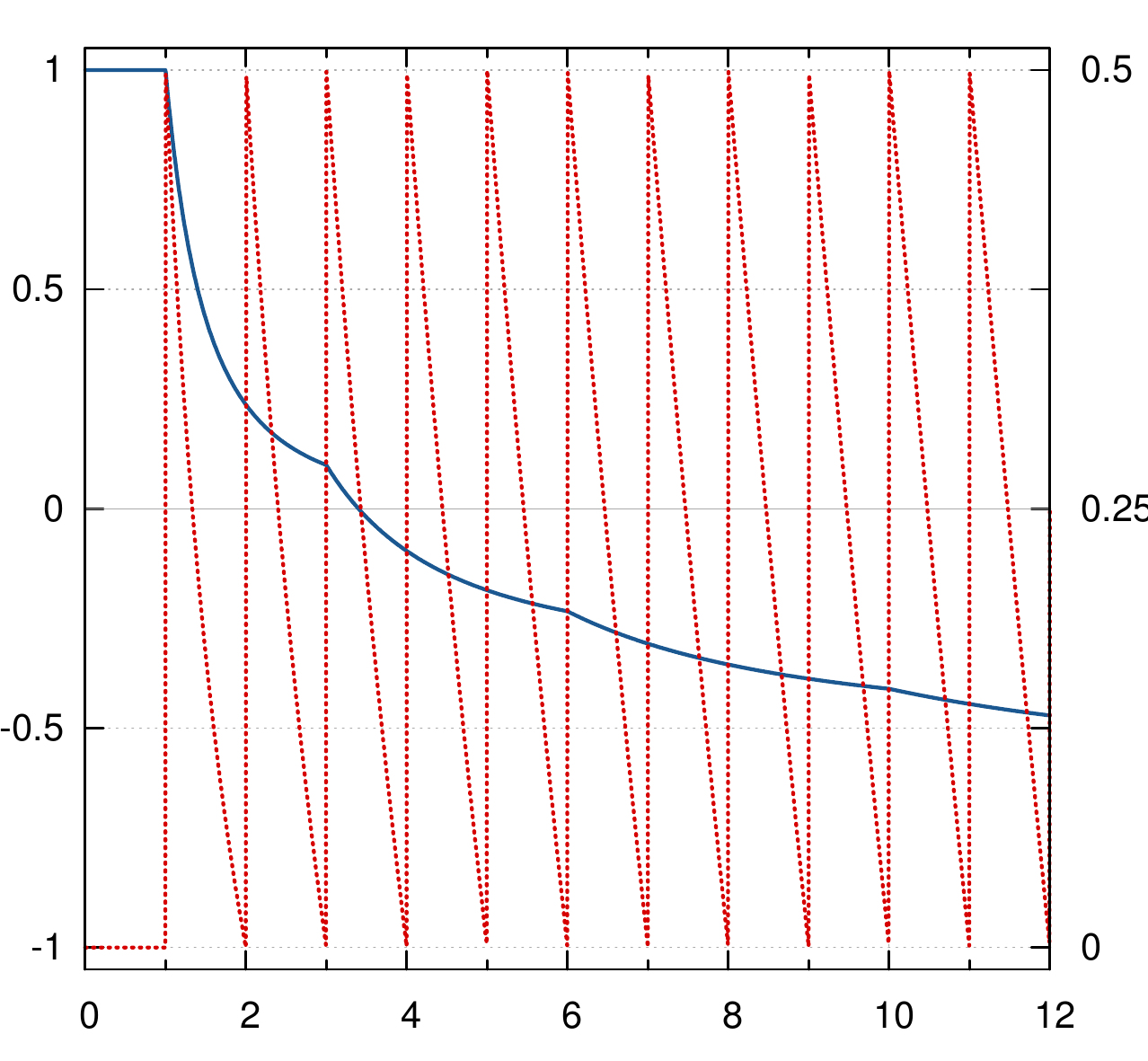}
 \caption{Dashed curve (red line, left scale): magnetization per particle $M/N\mu_B$ of a two-dimensional 
 electron gas at zero-temperature, as a function of $N\Phi_0/\Phi$, with $N$ the number of electrons and 
 $\Phi/\Phi_0$ the magnetic flux in units of the flux quantum $\Phi_0=h/e$. A discontinuous jump is observed each time a  
 Landau level  is completely filled. 
 Plain curve (blue line, right scale): dependence of the angular momentum $L_z/N^2\hbar$ on 
 $N\omega_{-}/\omega_{+}$ for a zero-temperature rotating trapped fermionic gas, in the limit of fast rotation.  
The angular momentum displays cusps at the discrete values $N\omega_{-}/\omega_{+}=p(p+1)/2=1,3,6,10,\cdots$, 
corresponding to the complete filling of an integer number of $n_{+}$ levels. 
At these points, the {\it derivative} of the angular momentum versus $N\omega_{-}/\omega_{+}$ displays discontinuous jumps.    
 This illustrates the similarities, but also the differences between the solid-state and ultra-cold fermionic gas contexts, 
 regarding the phenomenon of de Haas - van Alphen quantum oscillations.   
}
\label{fig:LandM}
\end{figure}

Let us consider the system at zero temperature with $N$ electrons, in the case where the first $p$ Landau levels are completely filled 
(each one with $\Phi/\Phi_0$ electrons) and the $(p+1)$-th one is partially filled with $N-p\Phi/\Phi_0$ electrons. 
The energy of the system 
$E=\sum_{n=0}^{p-1} (n+1/2)\hbar\omega_c\Phi/\Phi_0 + (N-p\Phi/\Phi_0) \hbar\omega_c (p+1/2)$ reads:   
\begin{equation}
\label{eq:electron_energy}
 E = \frac{\hbar\omega_c}{2}\frac{\Phi}{\Phi_0}p^2+\left( N-p\frac{\Phi}{\Phi_0}\right)(p+1/2)\hbar\omega_c\,.
\end{equation} 
Taking a derivative, one obtains the magnetization of the electron gas for $p\frac{\Phi}{\Phi_0} < N \leq (p+1)\frac{\Phi}{\Phi_0}$: 
\begin{equation}
 M = -\frac{\partial E}{\partial B} = \mu_B(2p+1)\left[p\frac{\Phi}{\Phi_0}-N \right]+\mu_B\, p \frac{\Phi}{\Phi_0}
\label{eq:magnetizations}
\end{equation}
with $\mu_B = \frac{\hbar e}{2m}$ the Bohr magneton. 
This expression shows that the magnetization jumps discontinuously by a macroscopic amount $2N\mu_B$ each time a new Landau level starts to 
be filled: when exactly $p$ Landau levels are filled the magnetization per electron is $M/N\mu_B=-1$ (Landau diamagnetism), and 
it jumps to $M/N\mu_B=+1$ as the $(p+1)$-th level starts to be filled, reaching smoothly $M/N\mu_B=-1$ again when this new level is completely 
filled. This behavior is depicted on Fig.~\ref{fig:LandM}, which displays $M/N\mu_B$ as a function of $N\Phi_0/\Phi$. 
These abrupt jumps are due to the macroscopic degeneracy of the Landau levels.
The magnetization per electron at $T=0$ is an approximately periodic function of $N\Phi_0/\Phi$, that is of particle number and {\it inverse} magnetic field. 

While this picture applies to 2DEGs, the electron dispersion along the $z$-axis is of course not negligible in most solid-state materials and 
must be taken into account (leading e.g. to the semi-classical picture of quantum oscillations in terms of extremal closed orbits on the 
Fermi surface).  
However, the simple picture above captures the essence of the de Haas - van Alphen effect.  

\subsection{Physical picture for a trapped rotating Fermi gas}

\subsubsection{Geometrical representation}
We now investigate the case of a rotating trapped atomic gas of fermions at $T=0$, focusing again for simplicity on a purely 
two-dimensional gas (strong transverse confinement $\omega_z\rightarrow\infty$).   
The main difference with the case of a 2DEG is that the single-particle spectrum is now that of a two-dimensional 
anisotropic harmonic oscillator, with {\it two different frequencies} $\omega_\pm=\omega_0\pm \omega$.
Hence, energy levels are non-degenerate, in contrast to the macroscopic degeneracy of the Landau levels typically present in the solids. 

\begin{figure}
 \centering
 \includegraphics[width=0.99\linewidth]{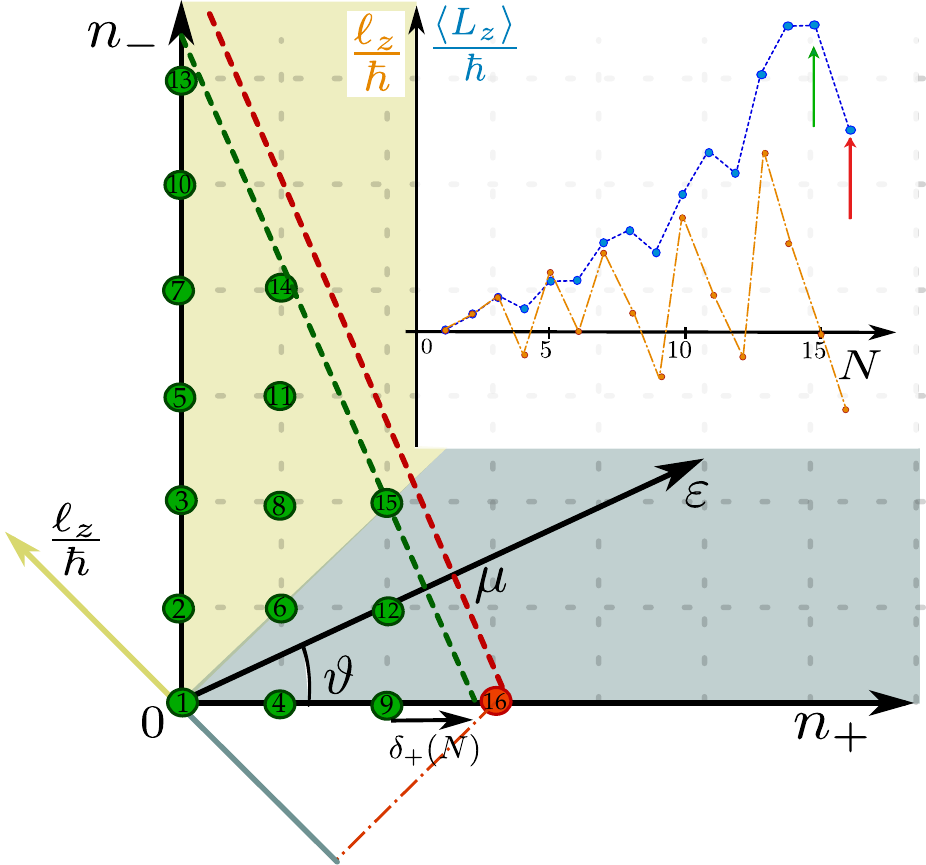}
 \caption{Sketch explaining the origin of the quantum oscillations for a two-dimensional gas at zero temperature (the case of low 
confinement is similar. In this case, each point of the graph will account for 
the particles that populate the $\omega_z$ levels allowed by the energy constraint). Each level is labeled by its integer coordinate $(n_+,n_-)$ and a projection onto the $\vartheta$ axis with slope $\omega_+/\omega_-$ gives the single particle energy $\varepsilon-\varepsilon_0$. The occupation of the levels is fixed by the chemical potential (green line). Green circles represent filled levels at zero temperature (satisfying the energy constraint), 
and the green situation depicted corresponds to the ground state of the system with N=15 particles. 
 The angular momentum corresponds to the projection onto the axis labeled by $\ell_z/\hbar$ as indicated by the dashed-dotted line. The two different colors for the axis and the corresponding shaded area indicate different sign of $\ell_z$.
The particles have been attributed a number (inside the green circle) corresponding to their order of appearance as the chemical 
potential is increased. Inset: The inset shows the evolution of the total angular momentum of the gas (in blue) with the number of particles, 
 and the angular momentum of each particle (orange). The arrows point the total angular momentum for $N=16$, and the contribution
from the supplementary particle added to the system by changing the chemical potential from the green to the red line.
}
\label{fig:QO_interp_picture}
\end{figure}

Fig.~\ref{fig:QO_interp_picture} helps to understand how this reflects on the angular momentum.  
This picture displays the single-particle states on a two-dimensional grid labeled by $(n_{+},n_{-})$. 
Because the energy levels (counted from the zero-point energy $\hbar\omega_0$) read 
$\hbar\omp\np+\hbar\omm\nm$, the value of the energy of a given state is obtained geometrically by projecting the 
representative point $(n_{+},n_{-})$ onto an axis tilted by an angle $\tan\theta=\omm/\omp$. Since 
each state can be filled by at most one fermion, the ground-state configuration for a given chemical 
potential (Fermi level) $\mu$ is obtained by occupying all states whose projection onto this axis is lower than $\mu$. 
The angular momentum of a single-particle state, on the other hand,  being given by $\ell_z/\hbar=n_- -n_+$ is obtained by projecting 
the point onto the anti diagonal of slope $-1$ (Fig.~\ref{fig:QO_interp_picture}). 
This means that particles exactly on the diagonal will not contribute to the angular momentum, whereas particles close to the vertical ($n_{-}$)  
and horizontal ($n_{+}$) axis will contribute most, with a positive and negative sign, respectively.  The evolution of $\ell_z$ is shown in the inset of Fig.~\ref{fig:QO_interp_picture}, ordering the particles with respect to the order in which states are filled.  

Increasing the chemical potential $\mu$ will lead to the filling of new states. In the situation sketched in Fig.~\ref{fig:QO_interp_picture} the difference between the two different chemical potentials corresponds to exactly adding one particle, namely particle number $16$. Since this particle has negative angular momentum, it will lead to a decrease of the total angular momentum $\langle L_z\rangle$ (cf.~Fig.~\ref{fig:QO_interp_picture}).

This simple picture shows very clearly the origin of the oscillations in the angular momentum. Roughly speaking, depending if during a shift of the chemical potential more particles are added in the upper diagonal (resp. lower) diagonal part, the angular momentum will increase (resp. decrease). For particles on the diagonal the angular momentum does not change.

\subsubsection{Simple picture at fast rotation}

For the case of fast rotation ($\omega\approx \omega_0$, so that $\omega_+ >>\omega_-$) this reasoning can be made more quantitative, and 
a simple calculation in the spirit of the one on 2DEGs above can be done.  
In this case, an analogue of the Landau levels are the successive $n_{+}$ levels. Further, the factor $\omega_{+}/\omega_{-}$ relates the number of $n_-$ levels with respect to the number of $n_+$ levels analogous to 
$\Phi/\Phi_0$, so that fast rotation corresponds to large magnetic fields.  
However, a crucial difference is that each $n_{+}$ level is not macroscopically degenerate, since $\omega_{-}$ is small but 
non-zero. 
The number of available $n_-$ states is nevertheless much larger than the number of available $n_+$ states. 
Therefore, we can assume a continuum of levels, in the $n_-$ direction, while treating the $n_+$ levels are discrete.   
Assume that at a certain chemical potential (particle number), the largest allowed $n_+$ quantum number is given by $p$. 
Then the largest allowed $n_-$ which corresponds to a certain fixed $n_+$ is given by $x_-(n_+)=\omega_+/\omega_- (p+\delta_+(N) -n_+)$, 
where $\delta_+ (N)= (N-\omega_+/\omega_-p(p+1)/2)/(\omega_+/\omega_-(p+1))$ is the distance of crossing 
of the chemical potential line and the $n_+$ axis from $p$, as depicted on Fig.~\ref{fig:QO_interp_picture}.

In this picture the angular momentum can be calculated as: 
\begin{eqnarray}
 \frac{\langle L_z \rangle}{\hbar}&& = \sum_{n_+=0}^{p}\int^{x_-(n_+)}_0 \left(n_--n_+\right)\\
&&= \frac{1}{2(1+p)}\left[N-\frac{p(1+p)}{2}\right]^2\nonumber\\
&&+\frac{p(p+1)}{24}\left[(p+2)\left(\frac{\hbar\omega_+}{\hbar\omega_-}+1\right)^2-2(p+1)\right]\,,
\label{eq:L_estimate_fastrot}
\end{eqnarray}
In the limit of very fast rotation $\omega_{+}\gg\omega_{-}$, this can be cast in the simpler form, valid for 
$p(p+1)/2 \leq \Neff \leq (p+1)(p+2)/2$: 
\begin{equation}
 \frac{\langle L_z \rangle}{N^2\hbar} = \frac{1}{2(p+1)}+\frac{p(p+1)(p+2)}{24\left[\frac{N\hbar\omega_-}{\hbar\omega_+}\right]^2}\,.
\end{equation}
which is to be compared to the form (\ref{eq:magnetizations}) for a 2DEG. In contrast to this case, it is easily 
checked that this formula is actually a {\it continuous function} of $\Neff$ (Fig.~\ref{fig:LandM}). 
At the special filling factors $\Neff=p(p+1)/2$ corresponding to starting to fill the  $p$-th level of the $\omega_{+}$ oscillator, 
$L_z$ does not display a discontinuous jump but rather a cusp, at which only the {\it derivative} of $L_z$ with respect to $\Neff$ 
has a discontinuity. This is clearly visible on Fig.~\ref{fig:LandM}.  
This difference in the behavior of $\langle L_z \rangle$ and the magnetization $M$ of an electron gas, which are analogous quantities, stems 
from the macroscopic degeneracy of the Landau level in the case of electrons. 
In the case of the rotating gas, the degeneracy is lifted due to the presence of the $\omega_{-}$ oscillator.\\

\subsubsection{Oscillations at slow rotation}

From this picture, and from the geometrical representation above, we expect that stronger oscillations are expected to occur in situations in which
the number of occupied $+$ and $-$ are comparable, i.e. for {\it slow rotation}. 
Furthermore, leaving the assumption of a two-dimensional system, we see that many particles can be added in the $\omega_z$ 
levels between successive $\omega_\pm$ levels if $\omega_z$ is small, thus leading to oscillations occurring for larger systems. 

\begin{figure}
 \centering
 \includegraphics[width=0.99\linewidth]{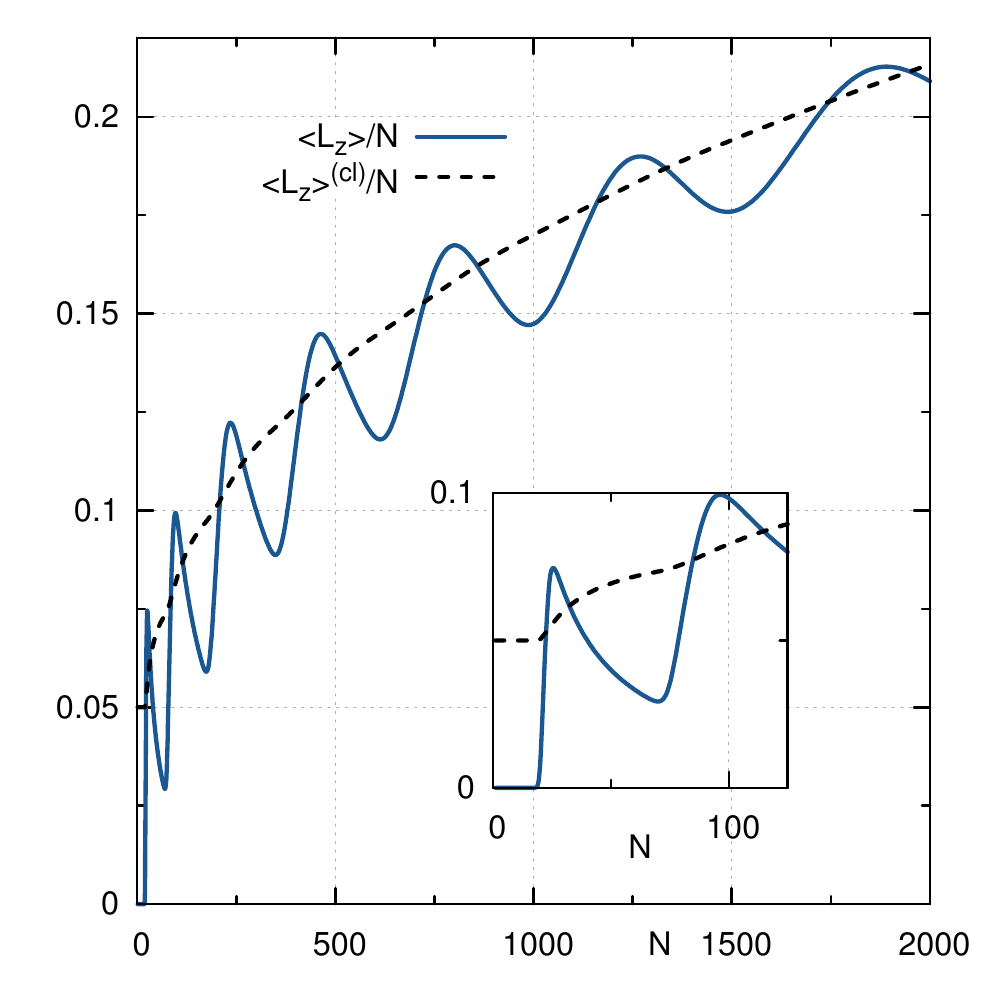}
 \caption{Angular momentum and its classical counterpart as a function of $N$, for $\alpha_z=0.05$, $T/T_F=0.01$ and 
$\alpha = 0.05$. The shell filling effect in the angular momentum is magnified due to the difference between $n_+$ and $n_-$ in its
expression, and results in an oscillating behavior. The inset shows the same two quantities, at low particle number. The shell filling effects
are visible in the classical angular momentum only at low particle number.
}
\label{fig:comparison_L}
\end{figure}

This simple scheme also explains why the oscillations are much less pronounced for the classical angular momentum (rigid-body moment of inertia). 
The $\omega_\pm$ levels contribute to this quantity with the same sign (Eq.~\eqref{eq:Lclass_result}). 
Hence, only shell-filling effects for very small systems are expected, as previously discussed for the radius of a mesoscopic gas 
confined to an anisotropic trap~\cite{PhysRevA.57.1253,PhysRevLett.85.4648,PhysRevA.67.053601,PhysRevA.58.2427,PhysRevA.55.4346}. 
These oscillations, e.g.~in the extension of the cloud are very hard to resolve experimentally and are typically only visible for very small atom numbers.
Fig.~\ref{fig:comparison_L} compares, for weak confinement, the effects in $L_z$ and in its classical analogue (extension of the gas). It is seen that in contrast to 
the latter, $L_z$ displays pronounced  oscillations as a function of the number of particles for rather large systems. 
This is because the angular momentum $\langle L_z \rangle$ involves the {\it difference} of $n_+$ and $n_-$. 
This also shows clearly that the main oscillatory contribution in~\eqref{eq:Ratiodef} stems from $\langle L_z \rangle^{(osc)}$. 

In this section, we have gained a helpful intuitive picture of the occurrence of oscillations in the angular momentum. The following 
section explores in more details the influence of the physical parameters of the system on the deviations of the angular momentum 
from its classical counterpart,  both analytically and numerically.

\section{Influence of physical parameters on quantum oscillations}
\label{sec:qcontributionsL}

In the forthcoming section, we will essentially focus on the oscillating contribution to the ratio of the quantum and classical angular momentum $R$. We discuss the impact of temperature, axial confinement and angular frequency on the amplitude and period of the oscillations, through the dimensionless
quantities $\frac{T}{T_F}$, $\alpha_z$ and $\alpha$, respectively. In an experiment, typically the atom number can be measured more easily than the chemical potential. Therefore, we monitor the evolution
of physical quantities as functions of the atom number rather than the chemical potential.
Since the amplitude of the oscillations will vary considerably with the system parameters we will point out the regimes in which the oscillations are best visible at reasonably large particle numbers.

  \subsection{The oscillation amplitude}

In this section, we focus on the influence of the system's parameters on the amplitude of the quantum oscillations.

  \subsubsection{Axial confinement and temperature}
\begin{figure}
 \centering
 \includegraphics[width=0.99\linewidth]{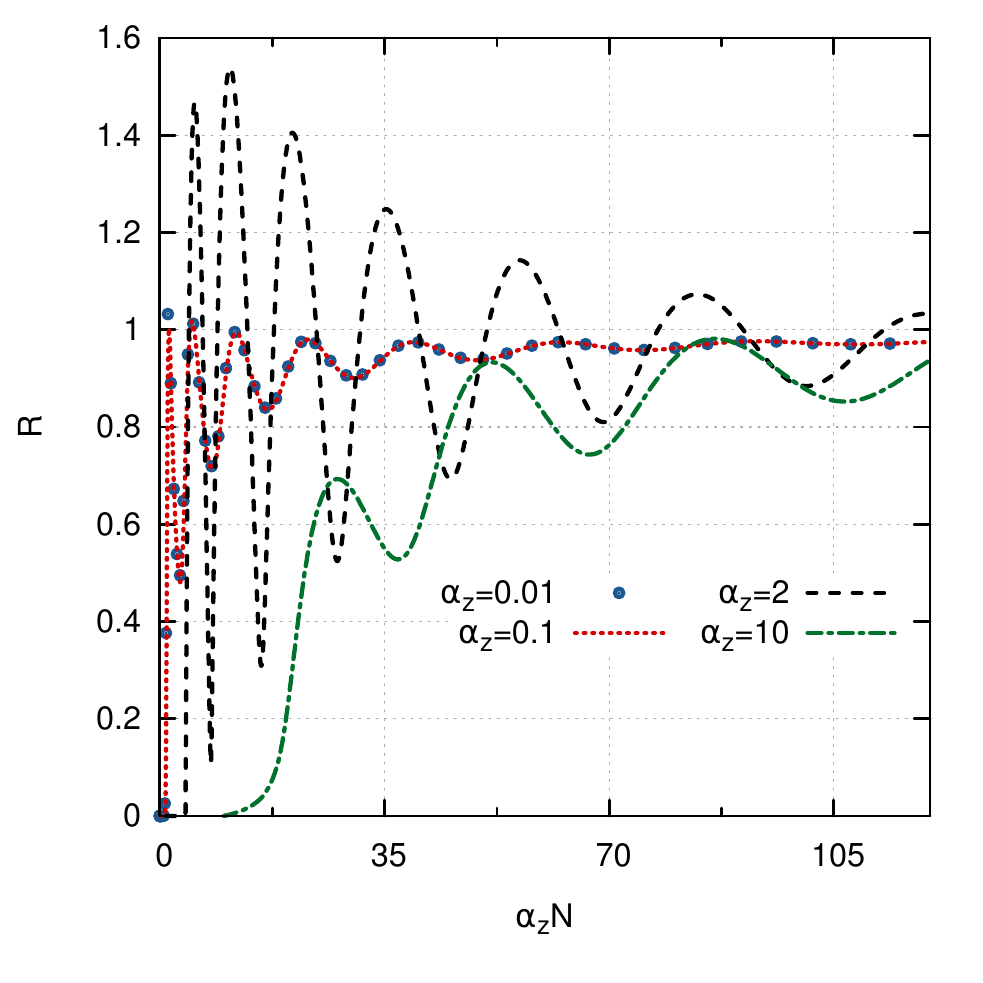}
 \caption{R vs $\alpha_z N$, at low temperature $T/T_F=0.03$ and $\alpha=0.05$, for $\alpha_z = 0.01$, $\alpha_z=0.02$, 
$\alpha_z=2$ and $\alpha_z=10$. Even though at low particle numbers, the amplitude of the oscillation is strongest for the intermediate aspect ratio $\alpha_z=2$, the oscillations remain visible for larger particle numbers in the weak confinement case. This indicates that a compromise between these two feature might be a favorable choice of
parameter to observe the oscillation of angular momentum.}
 \label{fig:az_comparison}
\end{figure}

Fig.~\ref{fig:az_comparison} compares the angular momentum ratio $R$ as a function of particle number for 
different values of the aspect ratio $\alpha_z$ of the trapping, and for identical values of angular frequency and temperature. For all parameters, the monotonous contribution leads to a rise of the ratio $R$ with increasing particle number towards the value one signaling a classical behavior of the angular momentum. On top of this rise, clear oscillations are evident at low particle number. The strong damping of the oscillations at large particle number is expected from the physical intuition given by Fig.~\ref{fig:QO_interp_picture}. The occupation of a new level becomes less important due to the large number of possible configurations. Also the denominator in \eqref{eq:Ratiodef} has larger values, and consequently a damping associated to the increasing particle number is visible. 

\begin{figure*}
 \centering
 \includegraphics[width=0.99\linewidth]{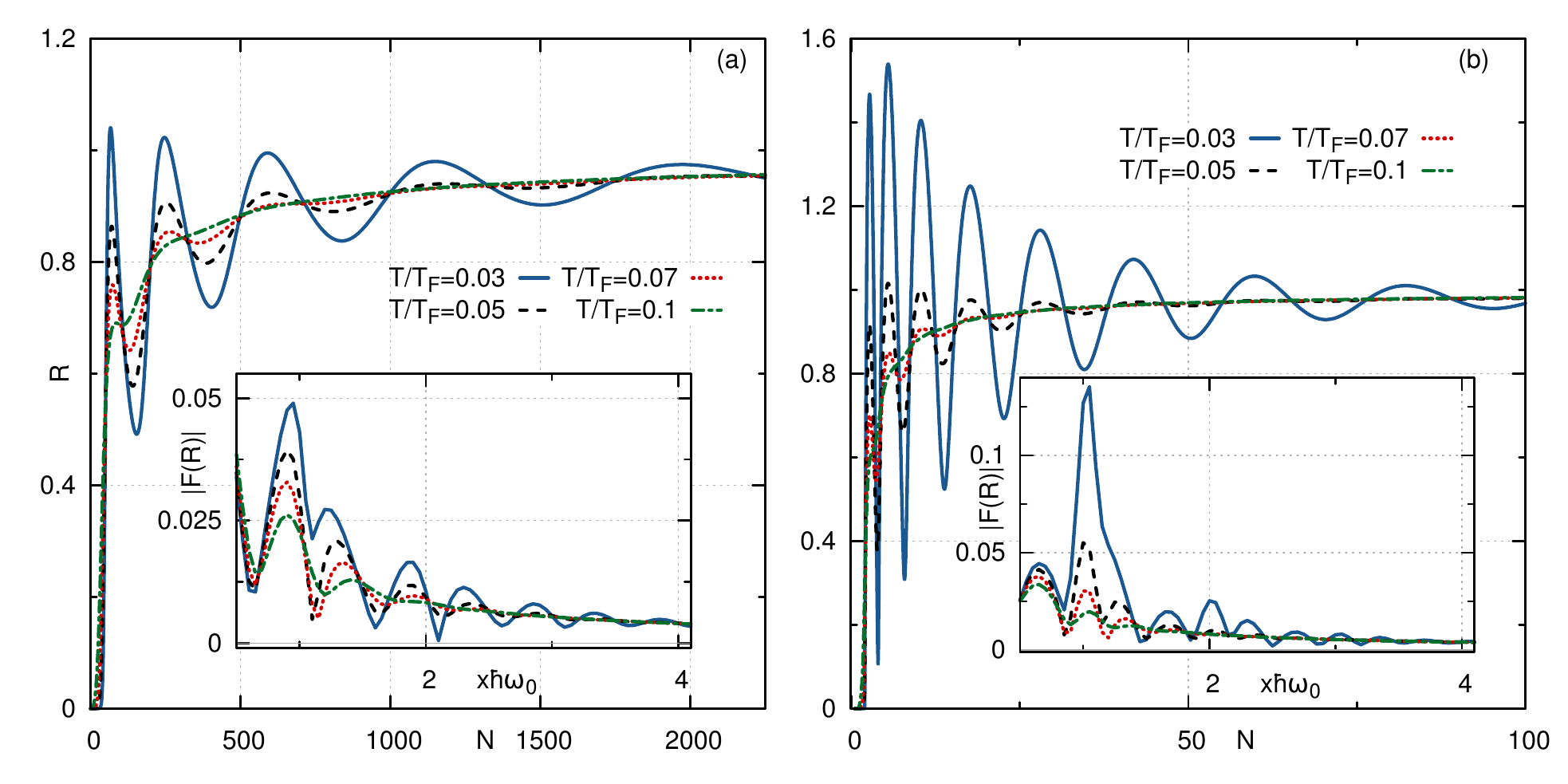}
 \caption{(a)- The ratio $R$ vs particle number, at low rotation frequency $\alpha = 0.05$ for $T/T_F = 0.03$, $T/T_F = 0.05$, $T/T_F = 0.07$ and $T/T_F = 0.1$. Panel (a) shows a weak aspect ratio of the confinement $\alpha_z=0.02$ and panel (b) an intermediate aspect ratio $\alpha_z=2$. In both cases, the low temperature curve displays the most pronounced
oscillations, and the oscillating behavior remains visible for larger samples. The stronger confinement leads to larger oscillation amplitudes, but the oscillations disappear for lower particle numbers. The insets shows the Fourier transform of $R$  wrt $\mu$ for the corresponding parameters. It shows that the amplitude of the oscillation modes decreases with the temperature, but the oscillation period remains unchanged. 
}
  \label{fig:comparison_T_lowaz}
\end{figure*}
The confinement along the $z$-direction controls how many levels are already at play before new $+,-$ levels contribute (cf. Fig.~\ref{fig:QO_interp_picture}). Therefore, the sharpness of the oscillations is strongly influenced by the aspect ratio of the trap as evident in Fig.~\ref{fig:az_comparison}. The largest amplitude of the oscillations is found at an almost isotropic confinement and low particle number $N$. However, at lower aspect ratio the oscillations survive till larger particle numbers, which might be a favorable situation to observe quantum oscillations in experiments.

\begin{figure} 
\centering
 \includegraphics[width=0.99\linewidth]{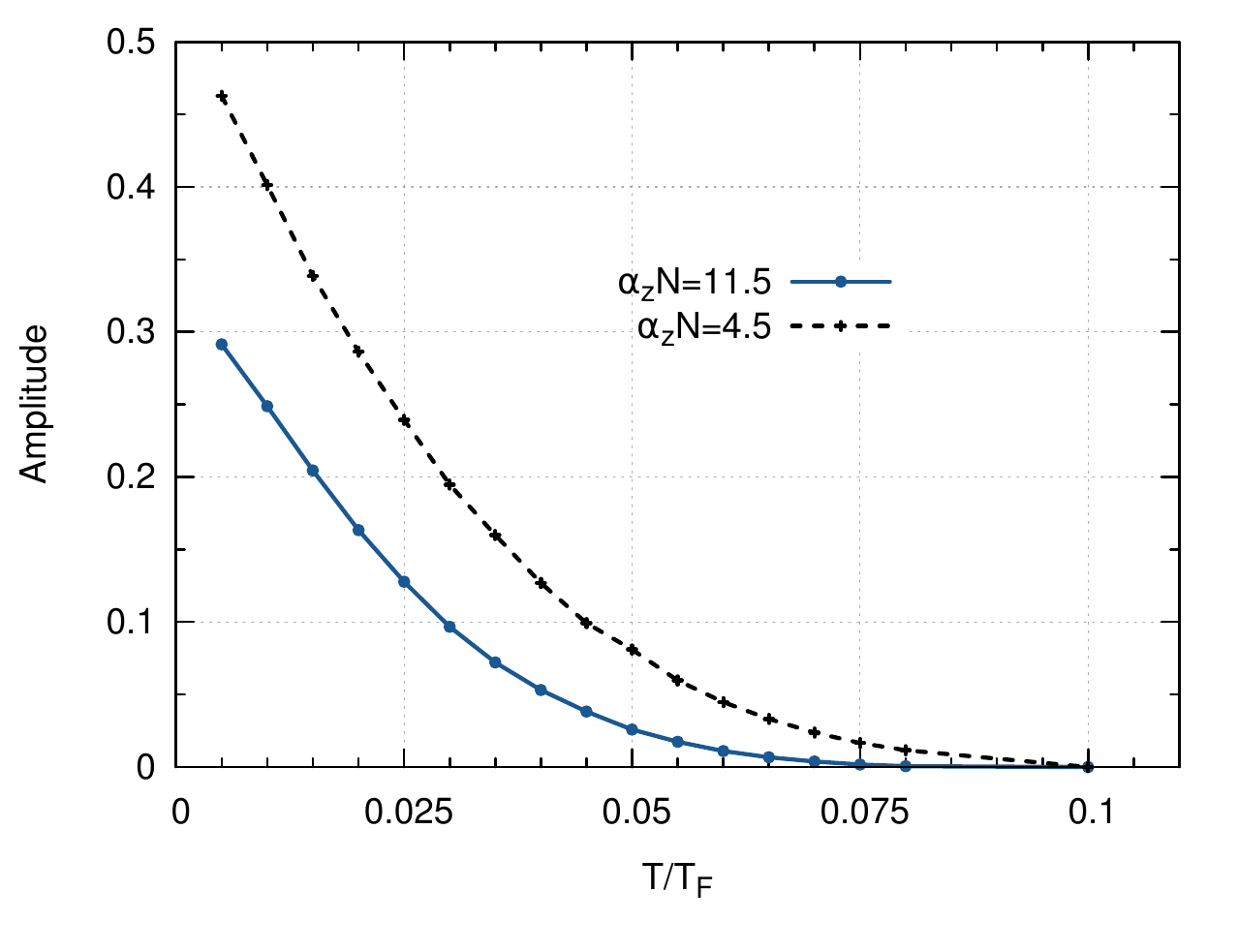}
 \caption{Amplitude of the oscillation of the ratio $R$ vs $T/T_F$, for $\alpha=0.05$ and for $\alpha_z N = 11.5$ (blue) and $\alpha_z N =4.5$ (black) with respect to the monotonous contribution~\eqref{eq:Rmon}.
As expected, at fixed confinement, the oscillation becomes less pronounced for bigger particle numbers. The strong damping effect due to the temperature
is clearly visible in both cases.}
  \label{fig:Temperature_effect}
\end{figure}
In the low $\alpha_z$ case, the oscillations
reach a {\it{universal}} regime, where the curves coincide as functions of $\alpha_z N$, as depicted on 
Fig.~\ref{fig:az_comparison}. This dependence on $\alpha_z N$ can be understood transforming the sum over $n_z$ in~\eqref{eq:thpot},
 into an integral, which gives for the thermodynamic potential :
\begin{equation}
\Omega = \frac{(k_BT)^2}{\hbar\omega_z}\sum_{n_+,n_-}\text{Li}_2[-e^{\beta(\mu-\varepsilon_0-\hbar\omega_+n_+-\hbar\omega_-n_-)}]\,.
\label{eq:Omega_lowaz}
\end{equation}
The only dependences on $\omega_z$ are in the prefactor and in $\varepsilon_0$. However, in the weak confinement regime,
$\omega_z \ll \omega_0$, and consequently the zero-point energy is dominated by $\varepsilon_0 \simeq \hbar\omega_0$. Taking the ratio $R$, the prefactor drops out in the expression~\eqref{eq:Ratiodef},
and one is left with a function of $\mu \propto (\alpha_z N)^{1/3}$ only. This universal regime strongly resembles the van Alphen oscillations in a solid. Due to its large extensions in a solid typically the electrons are assumed to move freely in the direction of the magnetic field.

The previous reasoning is valid at low temperature. At larger temperature, the sharpness of the Fermi surface is lost which will lead to less pronounced oscillations. Figures \ref{fig:comparison_T_lowaz}-(a) and \ref{fig:comparison_T_lowaz}-(b) compare the oscillations for different 
temperatures and fixed rotation, in the weak and strong confinement cases, respectively. A strong damping with increasing 
temperature of the amplitude of the oscillations is evident. 
The strong dependence of the oscillations on the sharpness of the Fermi surface plays
an important role in the observability of quantum oscillations, as known from the solid-state context~\cite{abrikosov1972introduction}, and
pointed out for cold atoms in~\cite{PhysRevA.57.1253,PhysRevA.67.053601,PhysRevA.58.2427,PhysRevA.55.4346}.

Fig.~\ref{fig:Temperature_effect} presents the amplitude of the oscillations as a function of temperature, at fixed rotation and 
axial confinement. A strong increase at low temperatures which seems almost linear occurs. 
This strong temperature dependence of the amplitude of the oscillating contribution can be understood from a Poisson summation analysis as presented in 
appendix~\ref{sec:Psums}, applied to $\langle L_z \rangle$ and $\langle L_z \rangle^{(cl)}$. By doing so, the temperature dependence only occurs in the sum in form of the expression $\left(\sinh{\left[\pi^2k_BT\left( \frac{k_+}{\hbar\omega_+}+\frac{k_-}{\hbar\omega_-}+\frac{k_z}{\hbar\omega_z}\right) \right]}\right)^{-1}$. This term causes the strong temperature damping of the oscillating
behavior of thermodynamic quantities. Let us note, that it also shows that the temperature has no influence on the oscillation period as we discuss below.

The strong dependence of the oscillation amplitude on temperature offers the possibility to use the oscillations as a thermometer in the 
low temperature regime. This low temperature regime ($T/T_F \lesssim 0.07$) is 
experimentally challenging, but has nevertheless been realized in a different context, with rotating \cite{MZwNature} or 
non rotating \cite{nascimbene-2010-463,Navon07052010} apparatus. A thermometer in this regime would be very important, since commonly used thermometry as the momentum distribution after time-of-flight imaging becomes less sensitive to temperature changes in this regime.  

\subsubsection{Rotation frequency}

In the solid-state case, quantum oscillations are most easily seen with high magnetic fields, which translate here into fast angular frequency.
 This statement finds its origin in the criterion that the cyclotron frequency $\omega_c = \frac{eB}{M}$ must be bigger than $k_BT$.\\
Fig.~\ref{fig:comp_rotation} depicts the evolution of $R$ with 
the particle number for different values of $\alpha$ at constant $T/T_F$ and axial confinement. In contrast to the solid state case 
the oscillations here are more pronounced when the rotation of the gas is slow, as already discussed qualitatively in section~\ref{sec:qualit_interp}. This can be related to the  occurrence of the two characteristic frequencies $\frac{1}{\hbar(\omega_0 \pm \omega)}$.  Indeed, $\alpha$ should not be too high such that $\hbar\omega_-$ remains bigger than $k_BT$.
This is expected from the Poisson summation analysis~: lower values of $\hbar\omega_-$ (corresponding to higher values of $\alpha$) 
tend to increase the temperature damping factor (see Appendix~\ref{sec:Psums}) and thereby decrease strongly the amplitude of the oscillations. 
Thus a strong dependence on the frequency $\omega$ is found.

Although $\alpha$ must be low compared to $1$, the rotation must not be too slow to be observable.

\subsection{(Pseudo)Period of oscillation}

In this subsection we focus on the effect of experimental parameters on the pseudo period of the oscillations. From the intuitive picture described in section~\ref{sec:qualit_interp}, one expects that the period of the oscillations does mainly depend on the energy level spacing given by $ \hbar \omega_{\pm,z}$. 

\begin{figure}
 \centering
 \includegraphics[width=0.99\linewidth]{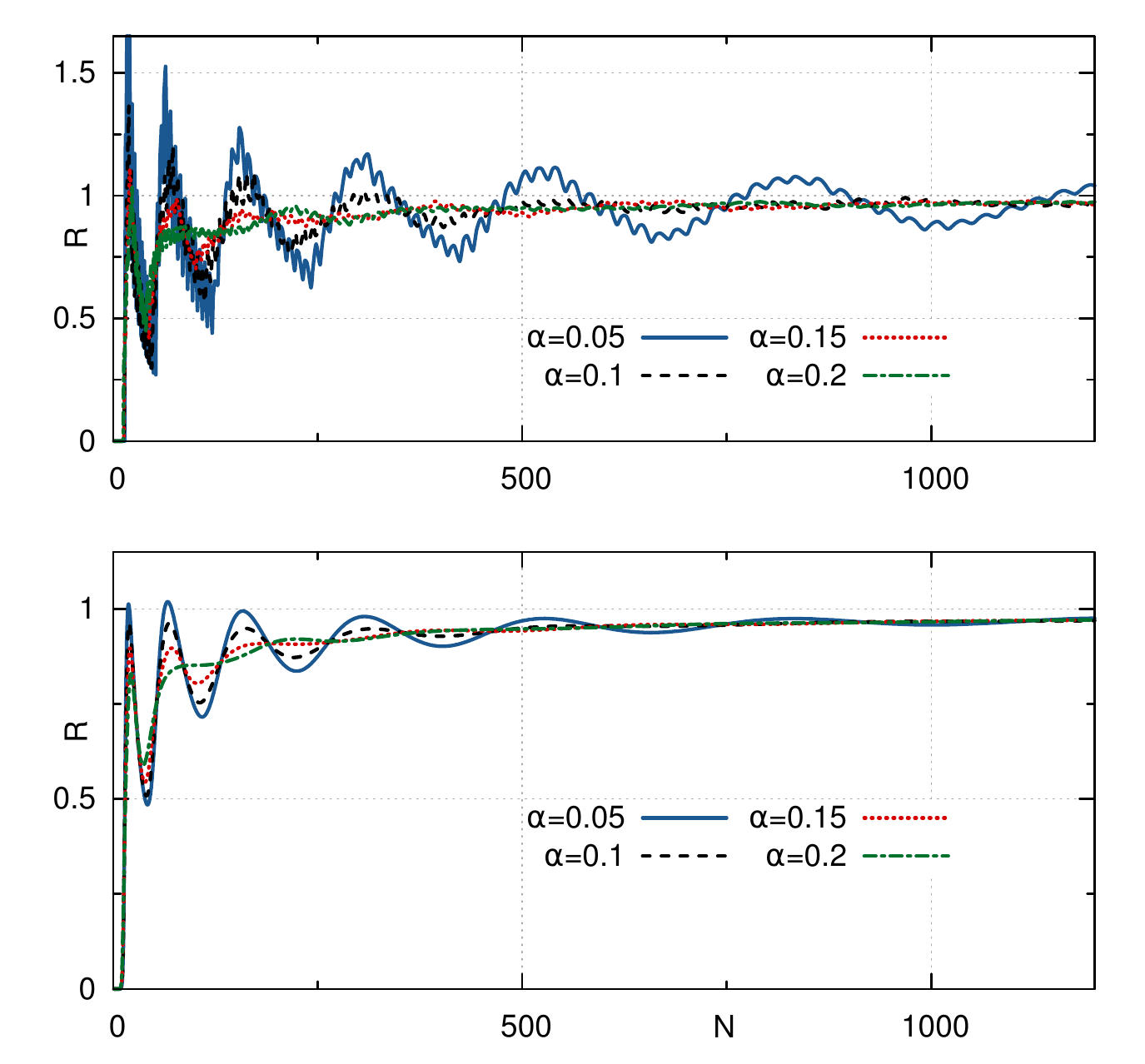}
 \caption{$R$ vs. particle number, at zero temperature (upper panel) and $T/T_F=0.03$ (lower panel) and weak trapping $\alpha_z=0.075$,
for several angular frequencies. Slow rotation favors 
the appearance of quantum oscillations. At zero temperature, fast oscillations due to the contribution of the $\omega_z$ levels appear.
}
 \label{fig:comp_rotation}
\end{figure}

\begin{figure}
  \centering
  \includegraphics[width=0.99\linewidth]{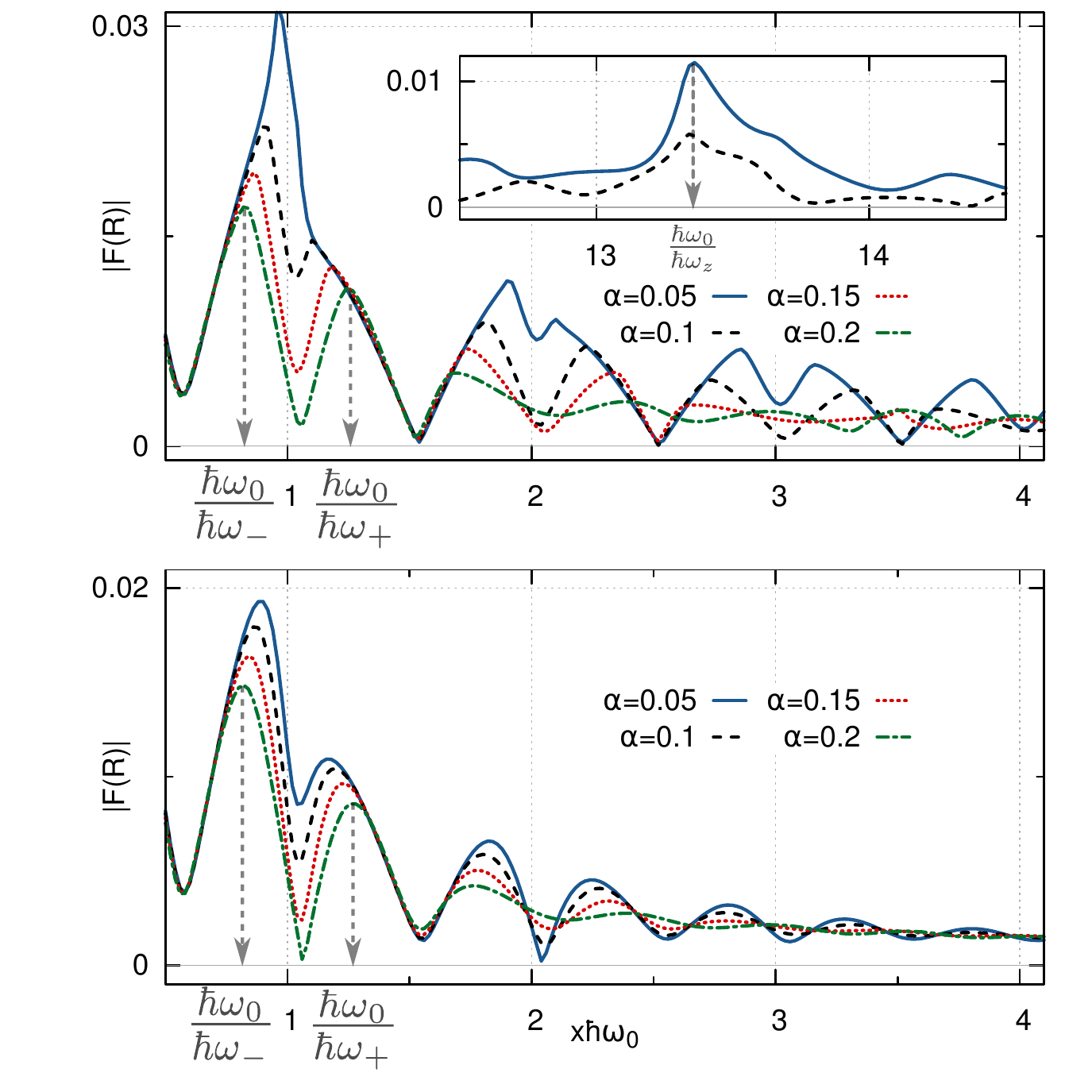}
  \caption{Fourier transform of R vs $x\hbar\omega_0$ at zero temperature (upper panel) and $T/T_F = 0.03$ (lower panel) , for 
fixed $\alpha_z = 0.075$. The successive harmonics of 
the fundamental frequencies are visible. The arrows indicate the position of the two characteristics 'frequencies' $\frac{1}{\hbar\omega_\pm}$.
Inset : Fourier transform of $R$ at zero temperature showing the peak corresponding to $\frac{1}{\hbar\omega_z}$, indicated by an arrow.
}
 \label{fig:R_vs_mu}
 \end{figure}

 The strong dependence on the two fundamental 'frequencies' $\frac{1}{\hbar\omega_+}$ and $\frac{1}{\hbar\omega_-}$  is confirmed 
by our numerical investigations reported in Fig.~\ref{fig:R_vs_mu}. This figure displays the modulus of the Fourier transform of the ratio $R$ \eqref{eq:Ratiodef} for several angular frequencies at zero temperature (upper panel) and for
$T/T_F=0.03$ (lower panel). The peaks that develop around integer values of $x\hbar\omega_0$ (the variable conjugated to $\mu$) show that
two fundamental 'frequencies' $\frac{1}{\hbar\omega_+}$ and $\frac{1}{\hbar\omega_-}$ appear (pointed out by arrows in Fig.~\ref{fig:R_vs_mu}). Increasing the frequency $\omega$ the peaks become more and more separated. 
Additionally to these main peaks, a peak at $\frac{1}{\hbar\omega_z}$ 
shows up here well separated from the other peaks (see inset in the upper panel of Fig.~\ref{fig:R_vs_mu}).
The interplay of these three different frequencies is also well seen in the quantum oscillation shown versus the particle number in Fig.~\ref{fig:comp_rotation}. The influence of $\omega_\pm$ can in particular be identified in Fig.~\ref{fig:comp_rotation} comparing $\alpha=0.1$ and $0.2$. A clear shift of the main oscillation period 
occurs. Additionally to the frequencies $\omega_\pm$, the trapping frequency $\omega_z$ causes smaller oscillations (Fig.~\ref{fig:comp_rotation}). Chosing a larger value of the aspect ratio of the trap, will influence considerably the oscillation period as shown in Fig.~\ref{fig:az_comparison}. 

Most of the detailed substructures of the oscillation discussed above are washed out at higher temperature, since the temperature causes the sharp 
peaks in the Fourier transform to broaden and only the main frequency determined by $\omega_0$ survives, as shown in Fig.~\ref{fig:comp_rotation} and~~\ref{fig:R_vs_mu}. 

One major difference compared to the solid state context is the existence of the three characteristic frequencies $\frac{1}{\hbar(\omega_0 \pm \omega)}$ and $
\frac{1}{\hbar\omega_z}$ in the oscillations, which is an effect of the radial and axial trapping potential, respectively. This occurrence is responsible for
an intrinsic limitation on angular frequency, which is forced to be lower than $\omega_0$ so that $\omega_-$
remains positive. This limitation is also at the origin
of the difficulties encountered to generate artificial magnetic fields strong enough to observe the quantum 
Hall effects in rotating gases \cite{RevModPhys.81.647,CooperReview,RevModPhys.80.885}. One possibility to circumvent this problem 
is to add a quartic term to the trapping potential, which has been implemented for example in \cite{PhysRevLett.92.050403}. 
This intrinsic limitation is also a motivation for the implementation of artificial gauge fields, which form the basis of
one of the protocols described in section~\ref{sec:Exp_imp}.


\section{Measurement of the angular momentum}
\label{sec:Exp_imp}

In section \ref{sec:qcontributionsL}, we have shown that the quantum oscillations of the angular momentum are best visible at low temperatures ($T/T_F \lesssim 0.07$) and slow rotation ($\alpha \lesssim 0.2$). A compromise between the particle number and the axial confinement has to be found. 
Even if hard to fulfill, these requirements are nevertheless on the reach of current experimental capacities. The height of the amplitude of the oscillations will be a very sensitive probe to the temperatures of the gas. 

\begin{figure}
 \centering
 \includegraphics[width=0.99\linewidth]{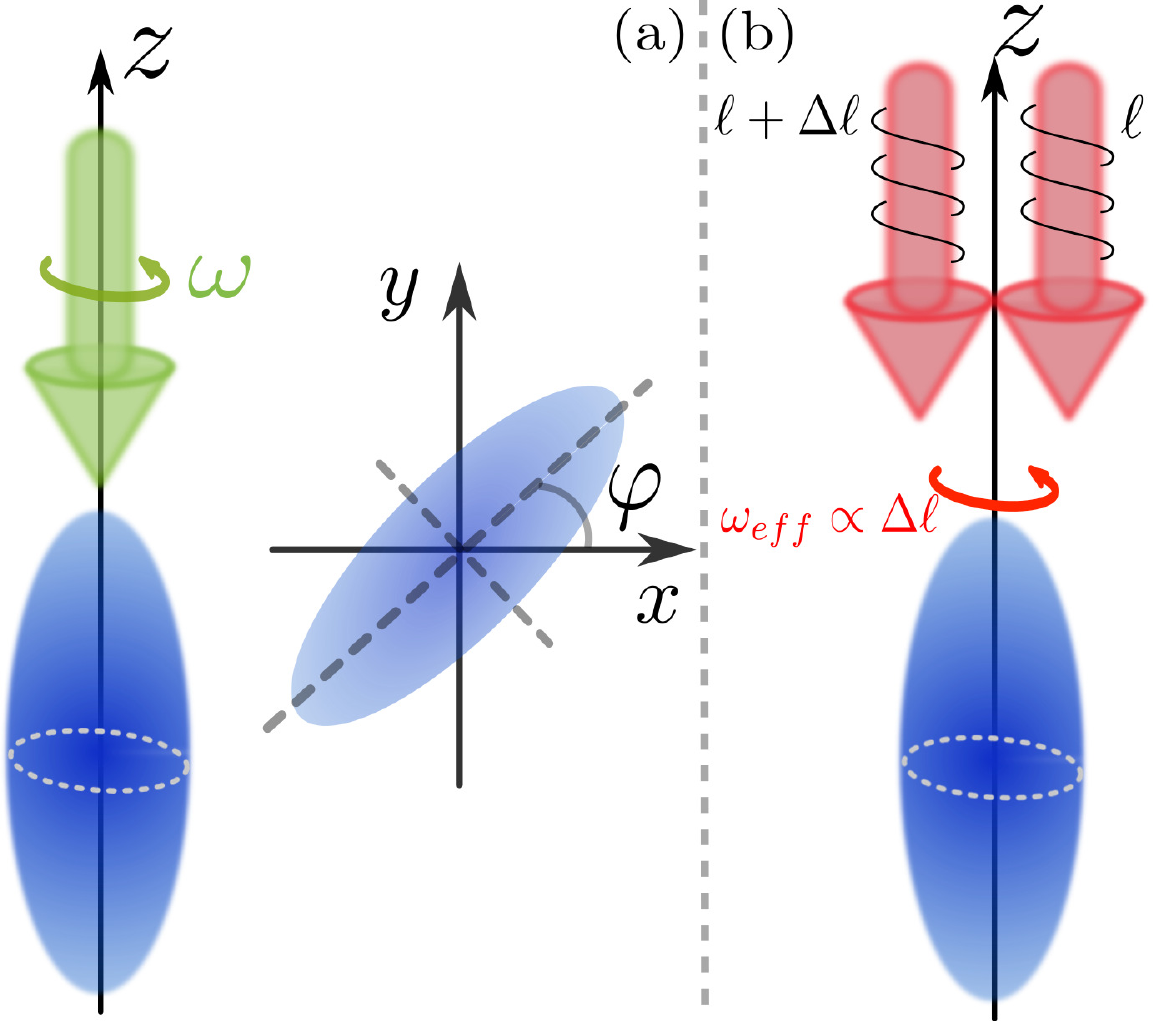}
 \caption{Sketch of the two different protocols for measuring the angular momentum of the fermion gas (see text for description): (a) Measurement
of the precession angle. The stirring beam (in green) is also used to create the anisotropic perturbation, which
provokes the appearance of quadrupole modes, and are responsible for a precession angle $\varphi$. Observing this angle provides a direct measurement of the angular
momentum of the gas . (b) Generation of
artificial rotation with lasers with spectroscopic measurement of the angular momentum. The two co propagating beams carry different angular momenta,
and the effective rotation is proportional to the angular momentum difference. The angular momentum is then measured by spectroscopy as suggested in~\cite{PhysRevLett.104.030401,PhysRevA.83.023610}.
}
  \label{fig:protocol}
\end{figure}

In order to measure $R$, one has to measure two observables : the square radius of the gas, and its angular momentum. 
Measurements of the density of the cloud and thus of its square radius can be performed using spatially resolved in-situ imaging techniques which are nowadays standard. For the observation of $\langle L_z \rangle$ two different experimental protocols (Fig.\ref{fig:protocol}) are useful. Since these are not commonly used,  we briefly summarize them in the following section~: 
\begin{itemize}
 \item Exciting the quadropole modes of the rotating gas and measuring the induced precession angle \cite{PhysRevLett.85.2223,PhysRevLett.84.806}, depicted in Fig.~\ref{fig:protocol}-(A)
 \item Measuring the occupation of internal states of the atoms which are connected to the angular momentum for light induced vector fields \cite{PhysRevLett.104.030401,PhysRevA.83.023610}, shown in Fig.~\ref{fig:protocol}-(B)
\end{itemize}

The first method has been successfully implemented with rotating Bose-Einstein condensates \cite{PhysRevLett.85.2223,PhysRevLett.84.806}. The underlying idea, as depicted in Fig.~\ref{fig:protocol}-(A), is to 
imprint an anisotropic perturbation of the form:
\begin{equation}
 V_{an} = \frac{V_0}{2} \left( x^2-y^2\right)\,,
\end{equation}
for a time $\delta t$ shorter than the rotation period $\omega^{-1}$~\cite{JDNotes}. This perturbation will cause an excitation of 
quadrupolar oscillation modes, which will lead to a precession of the cloud. The precession angle $\varphi$ of the main axis of the 
gas~(see Fig.~\ref{fig:protocol}-(A)) is related to the geometrical properties of the 
gas~:
\begin{equation}
 \tan{2\varphi} = \frac{2\langle xy \rangle}{\langle x^2\rangle-\langle y^2\rangle}\,.
\end{equation}
After the short excitation these properties change with time and can be computed perturbatively in the parameter $\epsilon = \frac{V_0\delta t}{\hbar}$:
\begin{align}
 \langle xy \rangle(t) & = \frac{\epsilon\hbar}{M^2}\langle L_z \rangle t^2 \\
 \langle x^2\rangle-\langle y^2\rangle(t) & = \frac{2\hbar\epsilon}{M}r_0^2t\,,
\end{align}
where $r_0^2$ is the radial size of the gas before the perturbation is applied. Assuming that $\varphi$ is small leads to the relation: 
\begin{equation}
 \varphi \simeq \frac{\langle L_z \rangle}{2Mr_0^2}t\,.
\end{equation}
Thus, following the evolution of this precession angle with time provides a direct measurement of $\langle L_z\rangle$.

The second technique to measure the angular momentum is described in Ref.~\cite{PhysRevLett.104.030401,PhysRevA.83.023610}, making use of artificial gauge
fields to generate an artificial slow rotation. 
More precisely, one has to shine on the gas two co propagating Laguerre-Gauss
beams carrying different angular momenta $\ell$ and $\ell+\Delta \ell$, as indicated in Fig.~\ref{fig:protocol}-(B). By two-photon Raman transition, the atoms become dressed by these beams and the dispersion of the lowest band is displaced by an effective shift to a non-zero angular momentum that is proportional to $\Delta \ell$. 
Since the generation of the vector field by this gauge field induces a coupling between the external (angular momentum) and internal (hyperfine) states of the atoms, the population of the internal states can be used to infer the angular momentum of the cloud.\\

Both these methods can be used as a thermometer of the fermionic cloud at low temperature, as suggested in section~\ref{sec:qcontributionsL}. By comparing directly the measured oscillation pattern with the exact expression derived 
from the thermodynamic potential~\eqref{eq:thpot}, knowing all the other parameters allows to determine the temperature of the gas. 
Furthermore, as Fig.~\ref{fig:Temperature_effect} has shown, the amplitude of the oscillatory phenomenon is expected to depend strongly 
on $T$, thus making the oscillation a very reliable temperature probe in the low temperature regime where current techniques fail.


\section{Summary and outlook}

In this article, we have investigated the phenomenon of quantum oscillations for a non-interacting Fermi gas subject 
to rotation or to an artificial magnetic field. 
We have considered the angular momentum of the gas, which is directly analogous to the magnetization 
in the solid-state context, and hence for which de Haas - van Alphen oscillations are expected as a function 
of e.g. atom number or chemical potential.

We have uncovered important differences between atomic gases and the solid-state context, especially 
in the limit of fast rotation $\omega \sim \omega_0$ for a two-dimensional system. These differences are due to the different degeneracies of 
the effective Landau levels. As a result, the pronounced periodic jumps in the magnetization per electron 
of a two-dimensional electron gas are replaced 
in the cold fermionic gas context by milder cusp-like singularities of the angular momentum versus atom number.

In contrast, we have shown that pronounced oscillations do show up in the opposite limit of slow rotation. 
The angular momentum is an observable which is particularly prone to oscillations, in contrast to 
the mean square radius of the gas for example (related to the rigid-body moment of inertia), which only 
displays small shell-filling effects for very small systems.  
We performed a detailed investigation of the dependence of these oscillations on experimentally relevant parameters. 
We concluded that they are  observable up to rather large atom numbers (several thousands) provided the harmonic 
confinement along the rotation axis is weak. 
The temperature must also be low compared to $\hbar\omega_{-}/k_B=\hbar(\omega_0-\omega)/k_B$, so 
that a significant effect for reasonably large systems is typically found only for $T/T_F\lesssim 0.1$.
In the solid-state context, quantum oscillations are a tool of choice for determining the Fermi surface of 
clean-enough materials~~\cite{abrikosov1972introduction,Ashcroft}. Such use is also possible in the context 
of atomic gases, but the technique can probably not compete with adiabatic band mapping~\cite{PhysRevLett.94.080403}. 
Quantum oscillations could, however, provide a useful method for in-situ thermometry in the regime of 
low $T/T_F$ where the amplitude of the oscillations depends strongly on temperature and where time-of-flight or in-situ measurements of the density profile are less sensitive. 

We proposed two experimental protocols which could be used to observe these quantum oscillations experimentally. 
These protocols elaborate on previous propositions to measure the angular momentum of the 
gas~\cite{PhysRevLett.85.2223,PhysRevLett.104.030401} inspired from the historic Andronikashviili 
experiment~\cite{AndronikaExperiment} on superfluid helium. 
There are several extensions of the present work that are worth considering. 
One is the effect of interactions on the quantum oscillations. Interactions and the associated finite lifetime of 
the quasiparticles will lead to a damping of the oscillations which will reduce their visibility~\cite{PhysRevA.58.2427}. 
At a phenomenological level, this can be accounted for by a Dingle-like factor, which in our case 
reads  $e^{-\gamma(n_{+}/\omega_{+}+n_{-}/\omega_{-})}$ 
in the oscillating contributions (assuming that the broadening due to interactions is the same for all energy levels). 
The characteristic damping $\gamma$ can be interpreted as the inverse lifetime of quasiparticles and 
its measurement (e.g. as a function of scattering length) could thus provide useful information on this lifetime. 

Another interesting direction is to investigate quantum oscillations in the presence of an optical lattice and 
an artificial gauge field, exploring in particular the possible signature of non-trivial topological states.

\section{Acknowledgements}
We warmly thank F. Chevy, J. Dalibard, S. Stringari and M. Zwierlein for fruitful discussions and experimental suggestions.
Financial support was provided by the ANR (project FAMOUS), the Triangle de la Physique (project CORSA),
the DARPA-OLE program, the SNSF under Division II and MaNEP.  

\appendix
\section{Poisson sums}
\label{sec:Psums}
Poisson summation is useful to compute sums of the form 
\begin{equation}
     \sum_{n = n_0}^{\infty} \phi (n)\,,
\end{equation}
and has been used by Landau to predict quantum oscillations for clean electronic systems \cite{Landau}. 
To describe the results presented in the main text, a sum over three variables $n_+,n_-,n_z$ is needed. 
In full generality, the result of this procedure is given by :
\begin{multline}
\label{eq:Psumresult}
 \sum_{\lbrace\nu\rbrace} \phi(\nu) = \int_{[0,+\infty[^3} d\vec{u} \phi(\vec{u}) + \\
 2\Re{\left\lbrace \sum_{\vec{k}\neq \vec{0}} \int_{[0,+\infty[^3} d\vec{u}
 \phi(\vec{u})e^{2i\pi \vec{k}\cdot\vec{u}} \right\rbrace}\,,
\end{multline}
where $\nu = (n_+,n_-,n_z)$ in the left hand side of eq.~\ref{eq:Psumresult} and $\vec{k}=(k_+,k_-,k_z)$ with $k_\pm,k_z$ positive integers.
The expression \eqref{eq:Psumresult} already displays the separation between monotonous and oscillatory contributions.
 
Let us apply this procedure to the average of a generic quantity $\Phi$ which reads
\begin{equation}
     \langle\Phi\rangle = \sum_{\lbrace\nu\rbrace}\frac{\Phi(\nu)}{1+e^{\beta(\varepsilon_\nu-\mu)}}.
\end{equation}

Using equation \eqref{eq:Psumresult}, this average can be separated into two contributions
\begin{equation}
     \label{eq:decomposition}
     \langle\Phi\rangle = \langle\Phi\rangle^{(no)} + \langle\Phi\rangle^{(osc)}\,.
\end{equation}
Here the monotonous part is given by
\begin{equation}
     \langle\Phi\rangle^{(no)}=\int_{[0,+\infty[^3}d\vec{u}\,\frac{\Phi(\vec{u})}{1+\exp{\left[\beta(\varepsilon_{\vec{u}}-\mu)\right]}}
\end{equation}
and the oscillating contribution by 
 \begin{multline}
      \langle\Phi\rangle^{(osc)} =2\Re\left\lbrace \sum_{\vec{k}\neq\vec{0}}\mathcal{F}(\vec{k})e^{2i\pi\left(\frac{k_+}{\hbar\omega_+}+\frac{k_-}{\hbar\omega_-}+\frac{k_z}{\hbar\omega_z}\right)\mu}\right\rbrace.
\end{multline}
 In the last expression, the quantity $\mathcal{F}(\vec{k})$ is given by :
\begin{widetext}
\begin{multline}
 \mathcal{F}(\vec{k}) = \frac{2}{3\hbar^3\omega_+\omega_-\omega_z}\int_{-\mu}^{+\infty}d\varepsilon\int_{-\infty}^{+\infty} dy'\int_{-\infty}^{+\infty} dz'
\frac{e^{2i\pi\left(\frac{k_+}{\hbar\omega_+}+\frac{k_-}{\hbar\omega_-}+\frac{k_z}{\hbar\omega_z}\right)\varepsilon}}{1+e^{\beta(\varepsilon+\varepsilon_0)}}e^{2i\pi\left(\frac{k_+}{\hbar\omega_+}-\frac{k_-}{\hbar\omega_-}\right)y'}e^{\frac{2i\pi}{3}\left(\frac{k_+}{\hbar\omega_+}+\frac{k_-}{\hbar\omega_-}-2\frac{k_z}{\hbar\omega_z}\right)z'}\\
\Phi\left(\frac{\varepsilon+\mu+y'/3+z'/9}{\hbar\omega_+/3},\frac{\varepsilon+\mu-y'/3+z'/9}{\hbar\omega_-/3},\frac{\varepsilon+\mu-2z'/9}{\hbar\omega_z/3}\right)\,,
\end{multline}
\end{widetext}
where the indices $k_\pm\,,k_z$ are all positive integers and $(k_+,k_-,k_z) \neq (0,0,0)$.\\
This analysis shows how the oscillating behavior arises in thermodynamical quantities such as the angular momentum studied here.
Furthermore, one sees that the temperature dependence of the oscillating part is contained in $\mathcal{F}$, which is the convolution
between the Fourier transform of $\Phi$ and the one of the occupation number, which behaves like 
$\left(\sinh{\left[\pi^2k_BT\left(\frac{k_+}{\hbar\omega_+}+\frac{k_-}{\hbar\omega_-}+\frac{k_z}{\hbar\omega_z}\right)\right]}\right)^{-1}$.
Thus, one can expect an exponential damping of the oscillating term due to temperature.\\

This computation also shows that the period of the oscillations are related to the inverse frequencies $\frac{1}{\hbar\omega_\pm}$ and $\frac{1}{\hbar\omega_z}$.
In our case, the axial trapping frequency is usually quite small, making the corresponding period very long, and thus not observable in realistic temperature conditions.
In most cases, the expressions for the oscillating part are quite complicated, and not very illuminating. 
Nevertheless, the decomposition~\eqref{eq:decomposition} provides rather simple expressions for the non oscillating contributions.


\section{Exact expressions for non oscillating contributions}
\label{sec:Analytics_NO}

The generalization of \eqref{eq:Psumresult} for several variables leads to the following expressions for the
non oscillating contributions, at any temperature :
\begin{align}
 & N = \frac{1}{(\beta\hbar\omega_0)^3(1-\alpha^2)\alpha_z}\mathcal{J}\\
 &\langle n_\pm \rangle^{(no)} = \frac{1}{(\beta\hbar\omega_0)^4(1-\alpha^2)\alpha_z(1\mp\alpha)}\mathcal{I}\\
 &\langle L_z \rangle^{(no)} = \frac{2\hbar\alpha}{(\beta\hbar\omega_0)^4(1-\alpha^2)^2\alpha_z}\mathcal{I}\\
&R_{mon} = \frac{\langle n_- \rangle^{(no)}-\langle n_+ \rangle^{(no)}}{\langle n_- \rangle^{(no)}+\langle n_+ \rangle^{(no)}+N}\\
\,\,\,{}&=\frac{2\mathcal{I}}{2\mathcal{I}+\beta\hbar\omega_0(1-\alpha^2)\mathcal{J}}\\
&\mathcal{I} = \frac{1}{6}\int_0^\infty du \frac{u^3}{1+e^{u-\xi}} = -\text{Li}_4[-e^{\beta(\mu-\varepsilon_0)}]\\
&\mathcal{J} = \frac{1}{2}\int_0^\infty du \frac{u^2}{1+e^{u-\xi}} = -\text{Li}_3[-e^{\beta(\mu-\varepsilon_0)}]\,,
\end{align}
with $\text{Li}_n$ the n-th polylogarithm function\cite{AStegun}.\\

These expressions show in particular that at large $\mu$, the populations of the radial harmonic oscillators are such that $\frac{\langle
n_+\rangle}{\langle n_- \rangle} = \frac{1-\alpha}{1+\alpha}$. Also, we see that the deviation of the ratio between the angular momentum of the gas
and its classical expression goes like $\frac{1}{\mu}$, thus like $N^{-1/3}$ \cite{PethickSmith} for macroscopic samples. The strictly zero
temperature result for the number of particles leads to $E_F = \hbar \omega_0 (6 \alpha_z N)^{1/3}$ \cite{PethickSmith}, as expected.


\section{Angular momentum of a classical gas}
\label{sec:Rotation_classical}

We consider a gas of classical particles interacting through the potential $V$ trapped in a potential $V_{trap}$.
The whole gas is rotating at constant angular frequency $\omega$ around the $z$-axis~:
\begin{equation}\label{eq:hamiltonian_classical}
 \mathcal{H} = \sum_i \frac{\vec{p}_i^2}{2M} +V_{trap}(\vec{r}_i)-L_{z,i}\omega
+\sum_{ij}V(|\vec{r}_i-\vec{r}_j|)\,.
\end{equation}
The average angular momentum of the gas is given by~: 
\begin{equation}
 \langle L_z \rangle = -\frac{\partial \text{Tr}\left( e^{-\beta\mathcal{H}}\right)}{\partial \omega}\,.
\end{equation}
One can rewrite the Hamiltonian \eqref{eq:hamiltonian_classical} as~:
\begin{align}
 \mathcal{H} & = \sum_i \frac{\left[ \vec{p}_i -M\vec{\omega}\times\vec{r}\right]^2}{2M} +V_{trap}(\vec{r}_i)
-\frac{1}{2}M\omega^2 (x^2+y^2) \nonumber \\
{} & +\sum_{ij}V(|\vec{r}_i-\vec{r}_j|)\,,
\end{align}
showing that, analogously to Bohr-van Leeuwen theorem (see eg \cite{blundell2001magnetism}), with an appropriate change of variables in the partition
function, one can immediately prove that the angular momentum is given by eq. \eqref{eq:Lclass_result}.
Furthermore, one also sees that the interaction potential $V$ does not play any role in deriving the expression
of the angular momentum, showing that the result~\eqref{eq:Lclass_result}
also holds for interacting particles.

\bibliography{QO} 

\begin{thebibliography}{37}%
\makeatletter
\providecommand \@ifxundefined [1]{%
 \@ifx{#1\undefined}
}%
\providecommand \@ifnum [1]{%
 \ifnum #1\expandafter \@firstoftwo
 \else \expandafter \@secondoftwo
 \fi
}%
\providecommand \@ifx [1]{%
 \ifx #1\expandafter \@firstoftwo
 \else \expandafter \@secondoftwo
 \fi
}%
\providecommand \natexlab [1]{#1}%
\providecommand \enquote  [1]{``#1''}%
\providecommand \bibnamefont  [1]{#1}%
\providecommand \bibfnamefont [1]{#1}%
\providecommand \citenamefont [1]{#1}%
\providecommand \href@noop [0]{\@secondoftwo}%
\providecommand \href [0]{\begingroup \@sanitize@url \@href}%
\providecommand \@href[1]{\@@startlink{#1}\@@href}%
\providecommand \@@href[1]{\endgroup#1\@@endlink}%
\providecommand \@sanitize@url [0]{\catcode `\\12\catcode `\$12\catcode
  `\&12\catcode `\#12\catcode `\^12\catcode `\_12\catcode `\%12\relax}%
\providecommand \@@startlink[1]{}%
\providecommand \@@endlink[0]{}%
\providecommand \url  [0]{\begingroup\@sanitize@url \@url }%
\providecommand \@url [1]{\endgroup\@href {#1}{\urlprefix }}%
\providecommand \urlprefix  [0]{URL }%
\providecommand \Eprint [0]{\href }%
\providecommand \doibase [0]{http://dx.doi.org/}%
\providecommand \selectlanguage [0]{\@gobble}%
\providecommand \bibinfo  [0]{\@secondoftwo}%
\providecommand \bibfield  [0]{\@secondoftwo}%
\providecommand \translation [1]{[#1]}%
\providecommand \BibitemOpen [0]{}%
\providecommand \bibitemStop [0]{}%
\providecommand \bibitemNoStop [0]{.\EOS\space}%
\providecommand \EOS [0]{\spacefactor3000\relax}%
\providecommand \BibitemShut  [1]{\csname bibitem#1\endcsname}%
\let\auto@bib@innerbib\@empty
\bibitem [{\citenamefont {Landau}(1930)}]{Landau}%
  \BibitemOpen
  \bibfield  {author} {\bibinfo {author} {\bibfnamefont {L.}~\bibnamefont
  {Landau}},\ }\href@noop {} {\bibfield  {journal} {\bibinfo  {journal} {Z.
  Phys}\ }\textbf {\bibinfo {volume} {64}},\ \bibinfo {pages} {629} (\bibinfo
  {year} {1930})}\BibitemShut {NoStop}%
\bibitem [{\citenamefont {de~Haas}\ and\ \citenamefont {van
  Alphen}(1930)}]{DHVA}%
  \BibitemOpen
  \bibfield  {author} {\bibinfo {author} {\bibfnamefont {W.}~\bibnamefont
  {de~Haas}}\ and\ \bibinfo {author} {\bibfnamefont {P.}~\bibnamefont {van
  Alphen}},\ }\href@noop {} {\bibfield  {journal} {\bibinfo  {journal} {Comm.
  Phys. Lab. Leiden}\ }\textbf {\bibinfo {volume} {212a}} (\bibinfo {year}
  {1930})}\BibitemShut {NoStop}%
\bibitem [{\citenamefont {Doiron-Leyraud}\ \emph {et~al.}(2007)\citenamefont
  {Doiron-Leyraud}, \citenamefont {Proust}, \citenamefont {LeBoeuf},
  \citenamefont {Levallois}, \citenamefont {Bonnemaison}, \citenamefont
  {Liang}, \citenamefont {Bonn}, \citenamefont {Hardy},\ and\ \citenamefont
  {Taillefer}}]{Naturecuprates}%
  \BibitemOpen
  \bibfield  {author} {\bibinfo {author} {\bibfnamefont {N.}~\bibnamefont
  {Doiron-Leyraud}}, \bibinfo {author} {\bibfnamefont {C.}~\bibnamefont
  {Proust}}, \bibinfo {author} {\bibfnamefont {D.}~\bibnamefont {LeBoeuf}},
  \bibinfo {author} {\bibfnamefont {J.}~\bibnamefont {Levallois}}, \bibinfo
  {author} {\bibfnamefont {J.-B.}\ \bibnamefont {Bonnemaison}}, \bibinfo
  {author} {\bibfnamefont {R.}~\bibnamefont {Liang}}, \bibinfo {author}
  {\bibfnamefont {D.~A.}\ \bibnamefont {Bonn}}, \bibinfo {author}
  {\bibfnamefont {W.~N.}\ \bibnamefont {Hardy}}, \ and\ \bibinfo {author}
  {\bibfnamefont {L.}~\bibnamefont {Taillefer}},\ }\href {\doibase
  10.1038/nature05872} {\bibfield  {journal} {\bibinfo  {journal} {Nature}\
  }\textbf {\bibinfo {volume} {447}},\ \bibinfo {pages} {565} (\bibinfo {year}
  {2007})}\BibitemShut {NoStop}%
\bibitem [{\citenamefont {Fetter}(2009)}]{RevModPhys.81.647}%
  \BibitemOpen
  \bibfield  {author} {\bibinfo {author} {\bibfnamefont {A.~L.}\ \bibnamefont
  {Fetter}},\ }\href {\doibase 10.1103/RevModPhys.81.647} {\bibfield  {journal}
  {\bibinfo  {journal} {Rev. Mod. Phys.}\ }\textbf {\bibinfo {volume} {81}},\
  \bibinfo {pages} {647} (\bibinfo {year} {2009})}\BibitemShut {NoStop}%
\bibitem [{\citenamefont {Cooper}(2008)}]{CooperReview}%
  \BibitemOpen
  \bibfield  {author} {\bibinfo {author} {\bibfnamefont {N.}~\bibnamefont
  {Cooper}},\ }\href {\doibase 10.1080/00018730802564122} {\bibfield  {journal}
  {\bibinfo  {journal} {Advances in Physics}\ }\textbf {\bibinfo {volume}
  {57}},\ \bibinfo {pages} {539} (\bibinfo {year} {2008})}\BibitemShut
  {NoStop}%
\bibitem [{\citenamefont {Bloch}\ \emph {et~al.}(2008)\citenamefont {Bloch},
  \citenamefont {Dalibard},\ and\ \citenamefont {Zwerger}}]{RevModPhys.80.885}%
  \BibitemOpen
  \bibfield  {author} {\bibinfo {author} {\bibfnamefont {I.}~\bibnamefont
  {Bloch}}, \bibinfo {author} {\bibfnamefont {J.}~\bibnamefont {Dalibard}}, \
  and\ \bibinfo {author} {\bibfnamefont {W.}~\bibnamefont {Zwerger}},\ }\href
  {\doibase 10.1103/RevModPhys.80.885} {\bibfield  {journal} {\bibinfo
  {journal} {Rev. Mod. Phys.}\ }\textbf {\bibinfo {volume} {80}},\ \bibinfo
  {pages} {885} (\bibinfo {year} {2008})}\BibitemShut {NoStop}%
\bibitem [{\citenamefont {Chevy}\ \emph {et~al.}(2000)\citenamefont {Chevy},
  \citenamefont {Madison},\ and\ \citenamefont
  {Dalibard}}]{PhysRevLett.85.2223}%
  \BibitemOpen
  \bibfield  {author} {\bibinfo {author} {\bibfnamefont {F.}~\bibnamefont
  {Chevy}}, \bibinfo {author} {\bibfnamefont {K.~W.}\ \bibnamefont {Madison}},
  \ and\ \bibinfo {author} {\bibfnamefont {J.}~\bibnamefont {Dalibard}},\
  }\href {\doibase 10.1103/PhysRevLett.85.2223} {\bibfield  {journal} {\bibinfo
   {journal} {Phys. Rev. Lett.}\ }\textbf {\bibinfo {volume} {85}},\ \bibinfo
  {pages} {2223} (\bibinfo {year} {2000})}\BibitemShut {NoStop}%
\bibitem [{\citenamefont {Zwierlein}\ \emph {et~al.}(2005)\citenamefont
  {Zwierlein}, \citenamefont {Abo-Shaeer}, \citenamefont {Schirotzek},
  \citenamefont {Schunck},\ and\ \citenamefont {Ketterle}}]{MZwNature}%
  \BibitemOpen
  \bibfield  {author} {\bibinfo {author} {\bibfnamefont {M.}~\bibnamefont
  {Zwierlein}}, \bibinfo {author} {\bibfnamefont {J.}~\bibnamefont
  {Abo-Shaeer}}, \bibinfo {author} {\bibfnamefont {A.}~\bibnamefont
  {Schirotzek}}, \bibinfo {author} {\bibfnamefont {C.}~\bibnamefont {Schunck}},
  \ and\ \bibinfo {author} {\bibfnamefont {W.}~\bibnamefont {Ketterle}},\
  }\href {\doibase 10.1038/nature03858} {\bibfield  {journal} {\bibinfo
  {journal} {Nature}\ }\textbf {\bibinfo {volume} {435}},\ \bibinfo {pages}
  {1047} (\bibinfo {year} {2005})}\BibitemShut {NoStop}%
\bibitem [{\citenamefont {Lin}\ \emph {et~al.}(2010)\citenamefont {Lin},
  \citenamefont {Compton}, \citenamefont {Jimenez-Garcia}, \citenamefont
  {Porto},\ and\ \citenamefont {Spielman}}]{SPNature}%
  \BibitemOpen
  \bibfield  {author} {\bibinfo {author} {\bibfnamefont {Y.-J.}\ \bibnamefont
  {Lin}}, \bibinfo {author} {\bibfnamefont {R.~L.}\ \bibnamefont {Compton}},
  \bibinfo {author} {\bibfnamefont {K.}~\bibnamefont {Jimenez-Garcia}},
  \bibinfo {author} {\bibfnamefont {J.~V.}\ \bibnamefont {Porto}}, \ and\
  \bibinfo {author} {\bibfnamefont {I.~B.}\ \bibnamefont {Spielman}},\ }\href
  {\doibase 10.1038/nature08609} {\bibfield  {journal} {\bibinfo  {journal}
  {Nature}\ }\textbf {\bibinfo {volume} {462}},\ \bibinfo {pages} {628}
  (\bibinfo {year} {2010})}\BibitemShut {NoStop}%
\bibitem [{\citenamefont {Dalibard}\ \emph {et~al.}(2011)\citenamefont
  {Dalibard}, \citenamefont {Gerbier}, \citenamefont {Juzeliunas},\ and\
  \citenamefont {Ohberg}}]{RevModPhys.83.1523}%
  \BibitemOpen
  \bibfield  {author} {\bibinfo {author} {\bibfnamefont {J.}~\bibnamefont
  {Dalibard}}, \bibinfo {author} {\bibfnamefont {F.}~\bibnamefont {Gerbier}},
  \bibinfo {author} {\bibfnamefont {G.}~\bibnamefont {Juzeliunas}}, \ and\
  \bibinfo {author} {\bibfnamefont {P.}~\bibnamefont {Ohberg}},\ }\href
  {\doibase 10.1103/RevModPhys.83.1523} {\bibfield  {journal} {\bibinfo
  {journal} {Rev. Mod. Phys.}\ }\textbf {\bibinfo {volume} {83}},\ \bibinfo
  {pages} {1523} (\bibinfo {year} {2011})}\BibitemShut {NoStop}%
\bibitem [{\citenamefont {Stringari}(1996)}]{PhysRevLett.76.1405}%
  \BibitemOpen
  \bibfield  {author} {\bibinfo {author} {\bibfnamefont {S.}~\bibnamefont
  {Stringari}},\ }\href {\doibase 10.1103/PhysRevLett.76.1405} {\bibfield
  {journal} {\bibinfo  {journal} {Phys. Rev. Lett.}\ }\textbf {\bibinfo
  {volume} {76}},\ \bibinfo {pages} {1405} (\bibinfo {year}
  {1996})}\BibitemShut {NoStop}%
\bibitem [{\citenamefont {Zambelli}\ and\ \citenamefont
  {Stringari}(1998)}]{PhysRevLett.81.1754}%
  \BibitemOpen
  \bibfield  {author} {\bibinfo {author} {\bibfnamefont {F.}~\bibnamefont
  {Zambelli}}\ and\ \bibinfo {author} {\bibfnamefont {S.}~\bibnamefont
  {Stringari}},\ }\href {\doibase 10.1103/PhysRevLett.81.1754} {\bibfield
  {journal} {\bibinfo  {journal} {Phys. Rev. Lett.}\ }\textbf {\bibinfo
  {volume} {81}},\ \bibinfo {pages} {1754} (\bibinfo {year}
  {1998})}\BibitemShut {NoStop}%
\bibitem [{\citenamefont {Haljan}\ \emph {et~al.}(2001)\citenamefont {Haljan},
  \citenamefont {Anderson}, \citenamefont {Coddington},\ and\ \citenamefont
  {Cornell}}]{PhysRevLett.86.2922}%
  \BibitemOpen
  \bibfield  {author} {\bibinfo {author} {\bibfnamefont {P.~C.}\ \bibnamefont
  {Haljan}}, \bibinfo {author} {\bibfnamefont {B.~P.}\ \bibnamefont
  {Anderson}}, \bibinfo {author} {\bibfnamefont {I.}~\bibnamefont
  {Coddington}}, \ and\ \bibinfo {author} {\bibfnamefont {E.~A.}\ \bibnamefont
  {Cornell}},\ }\href {\doibase 10.1103/PhysRevLett.86.2922} {\bibfield
  {journal} {\bibinfo  {journal} {Phys. Rev. Lett.}\ }\textbf {\bibinfo
  {volume} {86}},\ \bibinfo {pages} {2922} (\bibinfo {year}
  {2001})}\BibitemShut {NoStop}%
\bibitem [{\citenamefont {Leggett}(1970)}]{PhysRevLett.25.1543}%
  \BibitemOpen
  \bibfield  {author} {\bibinfo {author} {\bibfnamefont {A.~J.}\ \bibnamefont
  {Leggett}},\ }\href {\doibase 10.1103/PhysRevLett.25.1543} {\bibfield
  {journal} {\bibinfo  {journal} {Phys. Rev. Lett.}\ }\textbf {\bibinfo
  {volume} {25}},\ \bibinfo {pages} {1543} (\bibinfo {year}
  {1970})}\BibitemShut {NoStop}%
\bibitem [{\citenamefont {Leggett}(1998)}]{LeggetJSPHys}%
  \BibitemOpen
  \bibfield  {author} {\bibinfo {author} {\bibfnamefont {A.~J.}\ \bibnamefont
  {Leggett}},\ }\href@noop {} {\bibfield  {journal} {\bibinfo  {journal}
  {Journal of Statistical Physics}\ }\textbf {\bibinfo {volume} {93}},\
  \bibinfo {pages} {927} (\bibinfo {year} {1998})}\BibitemShut {NoStop}%
\bibitem [{\citenamefont {Abrikosov}(1972)}]{abrikosov1972introduction}%
  \BibitemOpen
  \bibfield  {author} {\bibinfo {author} {\bibfnamefont {A.}~\bibnamefont
  {Abrikosov}},\ }\href {http://books.google.fr/books?id=Jd9MbwAACAAJ} {\emph
  {\bibinfo {title} {{Introduction to the theory of normal metals}}}},\ {Solid
  State Physics: Supplement}\ (\bibinfo  {publisher} {Academic Press},\
  \bibinfo {year} {1972})\BibitemShut {NoStop}%
\bibitem [{\citenamefont {Cohen-Tannoudji}\ \emph {et~al.}(2006)\citenamefont
  {Cohen-Tannoudji}, \citenamefont {Diu}, \citenamefont {Laloe},\ and\
  \citenamefont {Diu}}]{Cohen}%
  \BibitemOpen
  \bibfield  {author} {\bibinfo {author} {\bibfnamefont {C.}~\bibnamefont
  {Cohen-Tannoudji}}, \bibinfo {author} {\bibfnamefont {B.}~\bibnamefont
  {Diu}}, \bibinfo {author} {\bibfnamefont {F.}~\bibnamefont {Laloe}}, \ and\
  \bibinfo {author} {\bibfnamefont {B.}~\bibnamefont {Diu}},\ }\href
  {http://www.worldcat.org/isbn/0471569526} {\emph {\bibinfo {title} {Quantum
  Mechanics (2 vol. set)}}}\ (\bibinfo  {publisher} {Wiley-Interscience},\
  \bibinfo {year} {2006})\BibitemShut {NoStop}%
\bibitem [{Note1()}]{Note1}%
  \BibitemOpen
  \bibinfo {note} {This monotonous deviation is also thought to be rather a
  quantum effect than a finite size one, since the classical expression for the
  angular momentum remains true even for a single classical
  particle}\BibitemShut {NoStop}%
\bibitem [{\citenamefont {Abramovitz}\ and\ \citenamefont
  {Stegun}(1972)}]{AStegun}%
  \BibitemOpen
  \bibfield  {author} {\bibinfo {author} {\bibfnamefont {M.}~\bibnamefont
  {Abramovitz}}\ and\ \bibinfo {author} {\bibfnamefont {I.}~\bibnamefont
  {Stegun}},\ }\href@noop {} {\emph {\bibinfo {title} {{Handbook of
  mathematical functions, {\it 10th edition}}}}},\ {}\ (\bibinfo  {publisher}
  {Dover},\ \bibinfo {year} {1972})\BibitemShut {NoStop}%
\bibitem [{\citenamefont {Schneider}\ and\ \citenamefont
  {Wallis}(1998)}]{PhysRevA.57.1253}%
  \BibitemOpen
  \bibfield  {author} {\bibinfo {author} {\bibfnamefont {J.}~\bibnamefont
  {Schneider}}\ and\ \bibinfo {author} {\bibfnamefont {H.}~\bibnamefont
  {Wallis}},\ }\href {\doibase 10.1103/PhysRevA.57.1253} {\bibfield  {journal}
  {\bibinfo  {journal} {Phys. Rev. A}\ }\textbf {\bibinfo {volume} {57}},\
  \bibinfo {pages} {1253} (\bibinfo {year} {1998})}\BibitemShut {NoStop}%
\bibitem [{\citenamefont {Ho}\ and\ \citenamefont
  {Ciobanu}(2000)}]{PhysRevLett.85.4648}%
  \BibitemOpen
  \bibfield  {author} {\bibinfo {author} {\bibfnamefont {T.-L.}\ \bibnamefont
  {Ho}}\ and\ \bibinfo {author} {\bibfnamefont {C.~V.}\ \bibnamefont
  {Ciobanu}},\ }\href {\doibase 10.1103/PhysRevLett.85.4648} {\bibfield
  {journal} {\bibinfo  {journal} {Phys. Rev. Lett.}\ }\textbf {\bibinfo
  {volume} {85}},\ \bibinfo {pages} {4648} (\bibinfo {year}
  {2000})}\BibitemShut {NoStop}%
\bibitem [{\citenamefont {Vignolo}\ and\ \citenamefont
  {Minguzzi}(2003)}]{PhysRevA.67.053601}%
  \BibitemOpen
  \bibfield  {author} {\bibinfo {author} {\bibfnamefont {P.}~\bibnamefont
  {Vignolo}}\ and\ \bibinfo {author} {\bibfnamefont {A.}~\bibnamefont
  {Minguzzi}},\ }\href {\doibase 10.1103/PhysRevA.67.053601} {\bibfield
  {journal} {\bibinfo  {journal} {Phys. Rev. A}\ }\textbf {\bibinfo {volume}
  {67}},\ \bibinfo {pages} {053601} (\bibinfo {year} {2003})}\BibitemShut
  {NoStop}%
\bibitem [{\citenamefont {Bruun}\ and\ \citenamefont
  {Burnett}(1998)}]{PhysRevA.58.2427}%
  \BibitemOpen
  \bibfield  {author} {\bibinfo {author} {\bibfnamefont {G.~M.}\ \bibnamefont
  {Bruun}}\ and\ \bibinfo {author} {\bibfnamefont {K.}~\bibnamefont
  {Burnett}},\ }\href {\doibase 10.1103/PhysRevA.58.2427} {\bibfield  {journal}
  {\bibinfo  {journal} {Phys. Rev. A}\ }\textbf {\bibinfo {volume} {58}},\
  \bibinfo {pages} {2427} (\bibinfo {year} {1998})}\BibitemShut {NoStop}%
\bibitem [{\citenamefont {Butts}\ and\ \citenamefont
  {Rokhsar}(1997)}]{PhysRevA.55.4346}%
  \BibitemOpen
  \bibfield  {author} {\bibinfo {author} {\bibfnamefont {D.~A.}\ \bibnamefont
  {Butts}}\ and\ \bibinfo {author} {\bibfnamefont {D.~S.}\ \bibnamefont
  {Rokhsar}},\ }\href {\doibase 10.1103/PhysRevA.55.4346} {\bibfield  {journal}
  {\bibinfo  {journal} {Phys. Rev. A}\ }\textbf {\bibinfo {volume} {55}},\
  \bibinfo {pages} {4346} (\bibinfo {year} {1997})}\BibitemShut {NoStop}%
\bibitem [{\citenamefont {Weisbuch}\ and\ \citenamefont
  {Vinter}(1991)}]{2DEGQO}%
  \BibitemOpen
  \bibfield  {author} {\bibinfo {author} {\bibfnamefont {C.}~\bibnamefont
  {Weisbuch}}\ and\ \bibinfo {author} {\bibfnamefont {B.}~\bibnamefont
  {Vinter}},\ }\href {http://nla.gov.au/nla.cat-vn1834745} {\emph {\bibinfo
  {title} {Quantum Semiconductor Structures: Fundamentals and Applications}}}\
  (\bibinfo  {publisher} {Academic Press},\ \bibinfo {year} {1991})\BibitemShut
  {NoStop}%
\bibitem [{\citenamefont {Nascimb{\`e}ne}\ \emph {et~al.}(2010)\citenamefont
  {Nascimb{\`e}ne}, \citenamefont {Navon}, \citenamefont {Jiang}, \citenamefont
  {Chevy},\ and\ \citenamefont {Salomon}}]{nascimbene-2010-463}%
  \BibitemOpen
  \bibfield  {author} {\bibinfo {author} {\bibfnamefont {S.}~\bibnamefont
  {Nascimb{\`e}ne}}, \bibinfo {author} {\bibfnamefont {N.}~\bibnamefont
  {Navon}}, \bibinfo {author} {\bibfnamefont {K.}~\bibnamefont {Jiang}},
  \bibinfo {author} {\bibfnamefont {F.}~\bibnamefont {Chevy}}, \ and\ \bibinfo
  {author} {\bibfnamefont {C.}~\bibnamefont {Salomon}},\ }\href
  {http://dx.doi.org/10.1038/nature08814} {\bibfield  {journal} {\bibinfo
  {journal} {Nature}\ }\textbf {\bibinfo {volume} {463}},\ \bibinfo {pages}
  {1057} (\bibinfo {year} {2010})}\BibitemShut {NoStop}%
\bibitem [{\citenamefont {Navon}\ \emph {et~al.}(2010)\citenamefont {Navon},
  \citenamefont {Nascimb{\`e}ne}, \citenamefont {Chevy},\ and\ \citenamefont
  {Salomon}}]{Navon07052010}%
  \BibitemOpen
  \bibfield  {author} {\bibinfo {author} {\bibfnamefont {N.}~\bibnamefont
  {Navon}}, \bibinfo {author} {\bibfnamefont {S.}~\bibnamefont
  {Nascimb{\`e}ne}}, \bibinfo {author} {\bibfnamefont {F.}~\bibnamefont
  {Chevy}}, \ and\ \bibinfo {author} {\bibfnamefont {C.}~\bibnamefont
  {Salomon}},\ }\href {\doibase 10.1126/science.1187582} {\bibfield  {journal}
  {\bibinfo  {journal} {Science}\ }\textbf {\bibinfo {volume} {328}},\ \bibinfo
  {pages} {729} (\bibinfo {year} {2010})}\BibitemShut {NoStop}%
\bibitem [{\citenamefont {Bretin}\ \emph {et~al.}(2004)\citenamefont {Bretin},
  \citenamefont {Stock}, \citenamefont {Seurin},\ and\ \citenamefont
  {Dalibard}}]{PhysRevLett.92.050403}%
  \BibitemOpen
  \bibfield  {author} {\bibinfo {author} {\bibfnamefont {V.}~\bibnamefont
  {Bretin}}, \bibinfo {author} {\bibfnamefont {S.}~\bibnamefont {Stock}},
  \bibinfo {author} {\bibfnamefont {Y.}~\bibnamefont {Seurin}}, \ and\ \bibinfo
  {author} {\bibfnamefont {J.}~\bibnamefont {Dalibard}},\ }\href {\doibase
  10.1103/PhysRevLett.92.050403} {\bibfield  {journal} {\bibinfo  {journal}
  {Phys. Rev. Lett.}\ }\textbf {\bibinfo {volume} {92}},\ \bibinfo {pages}
  {050403} (\bibinfo {year} {2004})}\BibitemShut {NoStop}%
\bibitem [{\citenamefont {Madison}\ \emph {et~al.}(2000)\citenamefont
  {Madison}, \citenamefont {Chevy}, \citenamefont {Wohlleben},\ and\
  \citenamefont {Dalibard}}]{PhysRevLett.84.806}%
  \BibitemOpen
  \bibfield  {author} {\bibinfo {author} {\bibfnamefont {K.~W.}\ \bibnamefont
  {Madison}}, \bibinfo {author} {\bibfnamefont {F.}~\bibnamefont {Chevy}},
  \bibinfo {author} {\bibfnamefont {W.}~\bibnamefont {Wohlleben}}, \ and\
  \bibinfo {author} {\bibfnamefont {J.}~\bibnamefont {Dalibard}},\ }\href
  {\doibase 10.1103/PhysRevLett.84.806} {\bibfield  {journal} {\bibinfo
  {journal} {Phys. Rev. Lett.}\ }\textbf {\bibinfo {volume} {84}},\ \bibinfo
  {pages} {806} (\bibinfo {year} {2000})}\BibitemShut {NoStop}%
\bibitem [{\citenamefont {Cooper}\ and\ \citenamefont
  {Hadzibabic}(2010)}]{PhysRevLett.104.030401}%
  \BibitemOpen
  \bibfield  {author} {\bibinfo {author} {\bibfnamefont {N.~R.}\ \bibnamefont
  {Cooper}}\ and\ \bibinfo {author} {\bibfnamefont {Z.}~\bibnamefont
  {Hadzibabic}},\ }\href {\doibase 10.1103/PhysRevLett.104.030401} {\bibfield
  {journal} {\bibinfo  {journal} {Phys. Rev. Lett.}\ }\textbf {\bibinfo
  {volume} {104}},\ \bibinfo {pages} {030401} (\bibinfo {year}
  {2010})}\BibitemShut {NoStop}%
\bibitem [{\citenamefont {John}\ \emph {et~al.}(2011)\citenamefont {John},
  \citenamefont {Hadzibabic},\ and\ \citenamefont
  {Cooper}}]{PhysRevA.83.023610}%
  \BibitemOpen
  \bibfield  {author} {\bibinfo {author} {\bibfnamefont {S.~T.}\ \bibnamefont
  {John}}, \bibinfo {author} {\bibfnamefont {Z.}~\bibnamefont {Hadzibabic}}, \
  and\ \bibinfo {author} {\bibfnamefont {N.~R.}\ \bibnamefont {Cooper}},\
  }\href {\doibase 10.1103/PhysRevA.83.023610} {\bibfield  {journal} {\bibinfo
  {journal} {Phys. Rev. A}\ }\textbf {\bibinfo {volume} {83}},\ \bibinfo
  {pages} {023610} (\bibinfo {year} {2011})}\BibitemShut {NoStop}%
\bibitem [{\citenamefont {Dalibard}(2000)}]{JDNotes}%
  \BibitemOpen
  \bibfield  {author} {\bibinfo {author} {\bibfnamefont {J.}~\bibnamefont
  {Dalibard}},\ }\href@noop {} {} (\bibinfo {year} {2000}),\ \bibinfo {note}
  {unpublished}\BibitemShut {NoStop}%
\bibitem [{\citenamefont {Ashcroft}\ and\ \citenamefont
  {Mermin}(1976)}]{Ashcroft}%
  \BibitemOpen
  \bibfield  {author} {\bibinfo {author} {\bibfnamefont {N.}~\bibnamefont
  {Ashcroft}}\ and\ \bibinfo {author} {\bibfnamefont {N.}~\bibnamefont
  {Mermin}},\ }\href@noop {} {\emph {\bibinfo {title} {{Solid State
  Physics}}}}\ (\bibinfo  {publisher} {Saunders College},\ \bibinfo {address}
  {Philadelphia},\ \bibinfo {year} {1976})\BibitemShut {NoStop}%
\bibitem [{\citenamefont {K{\"o}hl}\ \emph {et~al.}(2005)\citenamefont
  {K{\"o}hl}, \citenamefont {Moritz}, \citenamefont {St{\"o}ferle},
  \citenamefont {G{\"u}nter},\ and\ \citenamefont
  {Esslinger}}]{PhysRevLett.94.080403}%
  \BibitemOpen
  \bibfield  {author} {\bibinfo {author} {\bibfnamefont {M.}~\bibnamefont
  {K{\"o}hl}}, \bibinfo {author} {\bibfnamefont {H.}~\bibnamefont {Moritz}},
  \bibinfo {author} {\bibfnamefont {T.}~\bibnamefont {St{\"o}ferle}}, \bibinfo
  {author} {\bibfnamefont {K.}~\bibnamefont {G{\"u}nter}}, \ and\ \bibinfo
  {author} {\bibfnamefont {T.}~\bibnamefont {Esslinger}},\ }\href {\doibase
  10.1103/PhysRevLett.94.080403} {\bibfield  {journal} {\bibinfo  {journal}
  {Phys. Rev. Lett.}\ }\textbf {\bibinfo {volume} {94}},\ \bibinfo {pages}
  {080403} (\bibinfo {year} {2005})}\BibitemShut {NoStop}%
\bibitem [{\citenamefont {Andronikashvili}(1946)}]{AndronikaExperiment}%
  \BibitemOpen
  \bibfield  {author} {\bibinfo {author} {\bibfnamefont {E.}~\bibnamefont
  {Andronikashvili}},\ }\href@noop {} {\bibfield  {journal} {\bibinfo
  {journal} {J. Phys. USSR}\ }\textbf {\bibinfo {volume} {10}},\ \bibinfo
  {pages} {201} (\bibinfo {year} {1946})}\BibitemShut {NoStop}%
\bibitem [{\citenamefont {Pethick}\ and\ \citenamefont
  {Smith}(2008)}]{PethickSmith}%
  \BibitemOpen
  \bibfield  {author} {\bibinfo {author} {\bibfnamefont {C.}~\bibnamefont
  {Pethick}}\ and\ \bibinfo {author} {\bibfnamefont {H.}~\bibnamefont
  {Smith}},\ }\href@noop {} {\emph {\bibinfo {title} {{Bose-Einstein
  condensation in dilute gases}}}},\ {}\ (\bibinfo  {publisher} {Cambridge
  University Press},\ \bibinfo {year} {2008})\BibitemShut {NoStop}%
\bibitem [{\citenamefont {Blundell}(2001)}]{blundell2001magnetism}%
  \BibitemOpen
  \bibfield  {author} {\bibinfo {author} {\bibfnamefont {S.}~\bibnamefont
  {Blundell}},\ }\href {http://books.google.fr/books?id=V4V9QgAACAAJ} {\emph
  {\bibinfo {title} {{Magnetism in condensed matter}}}},\ {Oxford master}\
  (\bibinfo  {publisher} {Oxford University Press},\ \bibinfo {year}
  {2001})\BibitemShut {NoStop}%
\end{thebibliography}%

\end{document}